\documentclass[Conference]{IEEEtran}
\IEEEoverridecommandlockouts
\usepackage{cite}
\usepackage{amsmath,amssymb,amsfonts}
\usepackage{bm}
\usepackage{textcomp}

\usepackage[noend]{algorithmic}
\usepackage{algorithm}
\usepackage{graphicx}
\usepackage{epstopdf}
\usepackage{subfigure} 
\usepackage{booktabs}
\usepackage{multirow}
\usepackage{xcolor}
\usepackage{enumitem}
\usepackage{hyperref}
\hypersetup{hidelinks}

\def\BibTeX{{\rm B\kern-.05em{\sc i\kern-.025em b}\kern-.08em
		T\kern-.1667em\lower.7ex\hbox{E}\kern-.125emX}}
        
\begin{document}
\title{Optimal Packetization Towards Low Latency in Random Access Networks
}

\author{
\IEEEauthorblockN{Zihong Li, Anshan Yuan, and Xinghua Sun}
\IEEEauthorblockA{\\
    School of Electronics and Communication Engineering, Sun Yat-sen University, Shenzhen, China\\
	Email:\text{
lizh629@mail2.sysu.edu.cn, yuanansh@mail2.sysu.edu.cn, sunxinghua@mail.sysu.edu.cn}
    \thanks{This article is an extended version of a paper {presented} at  the IEEE 25th International Conference on Communication Technology (ICCT), Shenyang, China, October 2025\cite{MyConfPaper}.}
}\vspace{-1.2em}}

\maketitle
\begin{abstract}
As the demand for low-latency services grows, ensuring the delay performance of random access (RA) networks has become a priority. Existing studies on the queueing delay of the Aloha model universally treat packets as atomic transmission units, focusing on delay measured in time slots. However, the impact of packetization on queueing delay has been overlooked, particularly for the mean queueing delay measured in seconds. Here, packetization refers to the process of determining the number of bits assembled into a packet. This paper establishes the mathematical relationship between packetization and mean queueing delay in seconds for connection-free and connection-based Aloha schemes, and explores the optimal packetization to minimize the queueing delay. We identify the optimal packetization and its corresponding minimum  mean queueing delay via numerical methods, and analyze the influence of various network parameters. We further use simulations to investigate the impact of packetization on jitter of queueing delay. We then apply our analysis to re-evaluate the trade-off between the connection-free and connection-based schemes through the perspective of packetization. Furthermore, we apply the analysis to Random Access-Based Small Data Transmission (RA-SDT) in Non-Terrestrial
 Network (NTN) scenarios as a case study.
\end{abstract}


\section{Introduction} \label{SectionI}

Wireless communication technology is evolving from 5G to B5G and advancing toward 6G. A key driver of this evolution is the need to provide low-latency communication services, which is foundational for applications like industrial control, autonomous systems, and immersive media\cite{10529728, 10529954, 9464920}. Random access (RA) is an important technology for supporting these services and has been significantly enhanced across a series of 3GPP releases. A notable enhancement is the standardization of Random Access-Based Small Data Transmission (RA-SDT) in Release 17, a feature specifically designed for efficient delivery of small data payloads \cite{10230025}. Meanwhile, to realize the vision of global coverage, Non-Terrestrial Networks (NTNs) are being integrated into communication architectures \cite{9861699}. However, this integration poses a significant challenge, as the inherent large propagation delays in NTNs, in contrast to Terrestrial Networks (TNs), can profoundly affect the performance of RA protocols like RA-SDT.

Aloha is a fundamental RA model and underpins many practical RA protocols including RA-SDT \cite{10750858, 10154598}. In previous studies on Aloha's queueing delay performance, packets are uniformly treated as atomic units and queueing delay is typically evaluated in the abstract terms of time slots. However, this abstraction masks a fundamental optimization problem inherent in the packetization process, defined herein as the decision of how many bits to assemble into each packet. Aloha encompasses both connection-free scheme, where data packets directly contend for the channel, and connection-based scheme, where a short request reserves the channel before data transmission, and this packetization process affects the queueing delay performance of both. On one hand, for a given average bit arrival rate, creating smaller packets increases the packet arrival rate for both schemes, which intensifies channel contention and thus increases queueing delay. On the other hand, creating larger packets increases the time cost of each successful transmission; for the connection-free scheme, this extends the duration of each time slot, while for the connection-based scheme, it prolongs the channel occupancy time for a single transfer, during which other nodes are blocked from competing  for channel access to send requests. This in turn  increases queueing delay. Queueing delay measured in seconds provides a consistent metric for analyzing the relationship between packetization and queueing delay in both connection-free and connection-based schemes and offers greater practicality than queueing delay measured in time slots.

To address the above unexplored problem, this paper establishes a quantitative model for optimizing queueing delay in seconds through packetization. This model not only allows for the identification and analysis of the optimal packetization but also enables a more precise re-evaluation of the performance selection criteria between connection-free and connection-based schemes, and provides new perspectives for the performance analysis and optimization of RA-SDT in NTN scenarios.

\subsection{Delay Performance of Aloha}
Extensive studies have focused on the delay performance of Aloha from various perspectives and under different network conditions.

A number of studies have investigated the delay performance of Aloha with a single packet buffer \cite{1092814, 10.1145/322344.322345, 87169, 832998, 406614}. The delay analyzed in these works is essentially the access delay, defined as the total time from a packet becoming Head-of-Line (HOL) until its successful transmission. However, the access delay does not include the additional time spent by the packet waiting in the buffer to become a HOL packet, which is a common situation in real systems equipped with multiple packet buffers. In this case, queueing delay that includes the waiting time provides a more comprehensive performance metric.

To analyze the queueing delay, existing studies have proposed various analytical models, which can be mainly classified into Markov chain-based models and queueing theory-based models. In Markov chain-based studies, \cite{1102713} proposed an approach based on two Markov chains, which was subsequently refined into a coupled Markov chain model in \cite{1096749}. To reduce computational complexity, a refined approximate model based on two Markov chains was proposed in \cite{253331}. Additionally, an approximate model using an urn model analogy to calculate state transition probabilities was employed in \cite{318253}. In queueing theory-based studies, an individual node was modeled as an $M/G/1$ queue with exceptional service in \cite{332479}, and further modeled as an $M/G/1$ queue with multiple vacations in \cite{4967897} to capture the slotted boundary effects.

A group of studies has extended the analysis to incorporate physical layer characteristics, such as fading channels and the capture effect \cite{4155671, 1623486}, multipacket reception capabilities \cite{1459063}, and the availability of imperfect channel state information (CSI) \cite{4533941}. Furthermore, the effects of finite state Markov fading channels \cite{5336790} and OFDMA-based Aloha networks \cite{6373677} were also explored.

Exploring a different aspect, another group of studies has investigated various access mechanisms. These studies include the analysis of queue-aware transmission schemes \cite{8303680}, and on-demand sleep mechanisms \cite{10621020}. A key focus has been on connection-based Aloha scheme, for which analytical models were developed to evaluate its performance and characterize the trade-off between it and the connection-free scheme \cite{9244230, 10154598}. A significant contribution by \cite{10750858} established a unified analytical framework designed to evaluate and compare the mean queueing delay of Aloha and CSMA, which is adaptable to both connection-free and connection-based schemes, and various backoff schemes.

However, the above studies uniformly treat packets as whole entities of a fixed size and overlook the aforementioned fundamental optimization problem inherent in packetization.

\subsection{RA-SDT}
3GPP LTE Release 15 introduced Early Data Transmission (EDT), a random-access-based feature for small payloads in NB-IoT and LTE-M to improve battery life\cite{8412471}. \cite{7755762, 8885815, 10333294} have explored EDT or its similar precursor concepts from multiple angles. As an alternative, 3GPP LTE Release 16 introduced Preconfigured Uplink Resources (PUR), which, unlike the contention-based nature of EDT, relies on pre-configured radio resources to further reduce signaling overhead \cite{9139046}.

3GPP NR Release 17 further standardized the Small Data Transmission (SDT) feature. SDT is implemented in two main ways: the aforementioned contention-based RA-SDT and the reservation-based Configured Grant SDT (CG-SDT) \cite{10230025}. Specifically, RA-SDT supports both 2-step and 4-step implementations \cite{9448945}, which can be viewed as the connection-free Aloha and connection-based Aloha schemes through some approximation, respectively\cite{10750858}. \cite{10230025} innovatively integrated power-domain  Non-Orthogonal
Multiple Access (NOMA) with 2-step RA-SDT, leveraging reinforcement learning to enhance the transmission reliability for Reduced Capability (RedCap) devices at a low energy cost. The aforementioned studies \cite{10621020, 10154598, 10750858}, as well as another study about the energy efficiency of Aloha \cite{10606506}, have applied their respective theoretical analyses to RA-SDT in their case studies, providing useful insights. \cite{10621020} found that for 2-step RA-SDT, the on-demand sleep mechanism is clearly superior to the traditional duty-cycling sleep mechanism in terms of mean queueing delay and lifetime performance. \cite{10750858} found that introducing sensing and binary exponential backoff  into RA-SDT can reduce its mean queueing delay. The evaluation criteria for comparing the maximum data throughput and lifetime throughput performance of 2-step and 4-step RA-SDT were derived in \cite{10154598} and \cite{10606506}, respectively. \cite{10154598} also numerically analyzed their performance in the unsaturated region and mean queueing delay. Additionally, \cite{9709830, 9448945, 9771950, 10211103} have  evaluated and compared RA-SDT from various other perspectives.

It is worth noting that the above studies are almost exclusively focused on TN scenarios. However, the unique characteristics of NTNs, most notably the substantial propagation delay, profoundly alter the RA procedure. To our knowledge, an analysis of the queueing delay performance for RA-SDT in NTN scenarios, especially from a packetization perspective, remains an unexplored area.

\subsection{Our Contribution}
Motivated by the aforementioned limitations in the literature, this study moves beyond the conventional approach that treats packets as whole entities of a fixed size. Instead, we conduct an in-depth investigation into the relationship between packetization and mean queueing delay measured in seconds in Aloha networks. Building upon the unified analytical model for Aloha's mean queueing delay established in \cite{10750858}, our analysis encompasses both connection-free and connection-based schemes. It also re-evaluates their classic trade-off from the new perspective of packetization and applies the theoretical findings to RA-SDT in NTN scenarios as a case study.

{
The main contributions of this paper are summarized as follows:}
\begin{itemize}
    \item Based on the analytical framework from \cite{10750858}, we first derive the explicit relationship between packetization and the mean queueing delay (in seconds) for both connection-free and connection-based schemes. We also derive the expressions for the minimum packet size required to keep the network unsaturated, along with the corresponding feasibility conditions on network parameters.

    \item We employ numerical methods to analyze the relationship between packetization and mean queueing delay to identify the optimal packet size. This is followed by an investigation into the variation trends of both the optimal mean queueing delay and its corresponding optimal packet size with respect to different network parameters, which also examines the sensitivity differences between the two schemes. Furthermore, we use simulations to explore the similar relationship that exists between packetization and jitter of queueing delay, revealing that a degree of synergy exists between mean queueing delay and jitter of queueing delay performance for both schemes.

    \item Our above analysis is then used to re-evaluate the trade-off between connection-free and connection-based schemes. By characterizing three thresholds that divide the operational space into four distinct regions, we describe the complex relationship between the two schemes and explore how it is affected by network parameters. Finally, as a case study, we also apply our analysis to RA-SDT in NTN scenarios. We identify the scaling law relationship between the round trip time and both the optimal packet size and optimal delay. By comparing against an NR TN baseline, we also quantify the performance degradation in queueing delay and the variations in optimal packetization for both NR NTN and IoT NTN scenarios.
\end{itemize}

The remainder of this paper is organized as follows. Section \ref{sectionII} presents the system model and formulates the problem. In Section \ref{sectionIII}, we investigate the optimal packetization, analyze the impact of key network parameters, and further discuss the jitter of queueing delay. In Section \ref{sectionIV}, we leverage our analysis to re-evaluate the performance selection criteria between connection-free and connection-based schemes from the packetization perspective. In Section \ref{sectionV}, we apply our analysis to RA-SDT in NTN scenarios. Finally, Section \ref{sectionVI} concludes the paper.

\section{System Model and Problem Formulation}  \label{sectionII}
\subsection{System Model}

\begin{figure}[h] 
\centerline{\includegraphics[width=\linewidth]{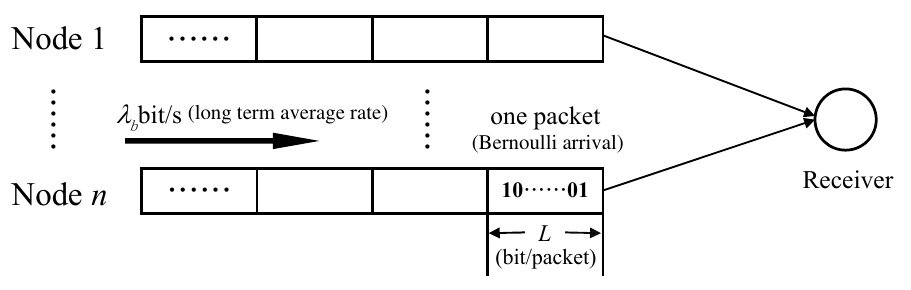}}
	\caption{Illustration of the system model.}
	\label{packetization}
\end{figure}

As illustrated in Fig.~\ref{packetization}, we consider a slotted Aloha network consisting of $n$ nodes and a single receiver. Each node generates a data bitstream at a long-term average rate of $\lambda_b$ bit/s. These bits are accumulated in a buffer and assembled into packets of size $L$ bits. The process of a new packet becoming fully formed and ready for transmission is modeled as a Bernoulli process, with packet arrivals occurring independently in each time slot. The generated packets are then transmitted to the receiver at an uplink data rate of $R$ bit/s.

\begin{figure}[htbp]
    \centering
    \subfigure[The connection-free Aloha.]{
        \label{connection-free}
        \begin{minipage}{\columnwidth}
            \centering

            \includegraphics[width=0.9\linewidth]{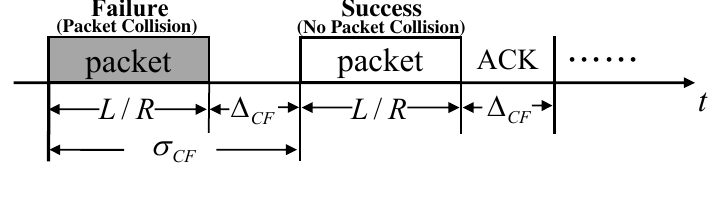}
        \end{minipage}
    }
    \vspace{1.5ex}
    \subfigure[The connection-based Aloha.]{
        \label{connection-based}
        \begin{minipage}{\columnwidth}
            \centering
            \includegraphics[width=\linewidth]{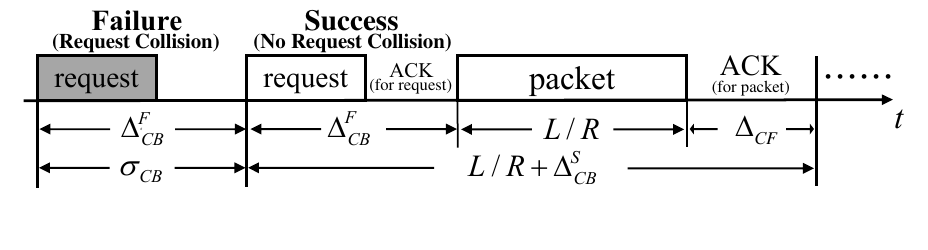}
        \end{minipage}
    }
    \caption{The time axis of (a) the connection-free and  (b) the connection-based Aloha.}
    \label{protocols}
\end{figure}

This paper investigates two schemes: connection-free and connection-based, whose time axes are illustrated in Fig.~\ref{protocols}. For both schemes, we assume the classic collision model where transmissions fail if multiple nodes transmit, whether data packets or requests, simultaneously, and succeed otherwise. Additionally, for brevity in figures and subsequent notations, we use CF and CB to denote the connection-free and connection-based schemes, respectively.

In the connection-free Aloha, the time axis is divided into time slots of  $\sigma_{CF} = L/R + \Delta_{CF}$ (s), consisting of the time to transmit the data packet $L/R$ (s) and the time for the acknowledgment (ACK) to confirm that the packet was received successfully $\Delta_{CF}$ (s).
Each node transmits its data packet at the beginning of a slot with a transmission probability $q$, and the entire data packet contends for the channel. In contrast, in connection-based Aloha, a node first sends a short request with transmission probability $q$ to reserve the channel, and it is this short request (not the data packet) that contends for the channel. The duration of this short request equals one slot length $\sigma_{CB}=\Delta_{CB}^{F}$ (s). We assume that the ACK duration used for confirming successful packet reception in the connection-based scheme  is equal to that in the connection-free scheme $\Delta_{CF}$ in this study. Upon a successful request, the channel is reserved for the corresponding node to transmit its data packet during the subsequent $L/R+\Delta_{CF}$ (s), while other nodes are blocked from sending requests in this period. The entire successful transmission lasts $L/R+\Delta_{CB}^{S}= L/R + \Delta_{{CB}}^{F} + \Delta_{{CF}}$ (s).

\subsection{Problem Formulation}
{
The packet arrival rate $\lambda$, defined as the expected number of packets generated per slot for each node, is derived from the bit arrival rate $\lambda_b$ and the slot duration $\sigma$. Specifically, the relationship is given by $\lambda = \frac{\lambda_b}{L}\sigma$.}

{
For the connection-free scheme, substituting the slot duration $\sigma_{CF} = L/R + \Delta_{CF}$ yields:
\begin{equation}
\lambda_{CF} = \frac{\lambda_b}{L} \left( \frac{L}{R} + \Delta_{CF} \right) = \lambda_b \left( \frac{1}{R} + \frac{\Delta_{CF}}{L} \right).
\end{equation}
}

{
For the connection-based scheme, substituting the request slot duration $\sigma_{CB} = \Delta_{CB}^{F}$ yields:
\begin{equation}
\lambda_{CB} = \frac{\lambda_b}{L} \Delta_{CB}^{F}.
\end{equation}
}

The mean queueing delay  measured in seconds, $\overline{T}$, is the product of the mean queueing delay  measured in time slots  $\overline{T}_{ts}$ and the slot duration $\sigma$. Modeling the above system as a Geo/G/1 queueing system, $\overline{T}_{ts}$ can be expressed as \cite{Takagi1993}:
\begin{equation}
    \overline{T}_{ts} = \frac{\lambda \overline{D}^2 - \lambda \overline{D}}{2(1 - \lambda \overline{D})} + \overline{D},
\end{equation}
where $\overline{D}$ and $\overline{D}^2$ are the first and second moments of the service time, provided the queue is unsaturated.

The service time moments for each scheme can be obtained by specializing the general analytical framework established in \cite{10750858} to our  model. For the connection-free scheme, this yields:
\begin{equation}
    \overline{D}_{CF} = \frac{1}{q \cdot e^{\mathbb{W}_0(-n\lambda_{CF})}},
\end{equation}
\begin{equation}
    \overline{D}_{CF}^2 = \frac{2}{q^2 \cdot e^{2\mathbb{W}_0(-n\lambda_{CF})}} - \overline{D}_{CF},
\end{equation}
{where $\mathbb{W}_0(\cdot)$ denotes the principal branch of the Lambert W function \cite{corless1996lambert}.}

For the connection-based scheme, this yields:
\begin{equation}
    \overline{D}_{CB} = \tau_T - 1 + \frac{1}{p \tilde{\alpha} q},
\end{equation}
\begin{equation}
    \overline{D}_{CB}^2 = \frac{2}{p\tilde \alpha q}(\frac{1}{p\tilde \alpha q}+\tau_T-2)+(\tau_T-1)(\tau_T-2) +  \overline{D}_{CB},
\end{equation}
where $\tau_T = (L/R + \Delta_{CB}^{S}) / \Delta_{CB}^{F}$ is the normalized successful cycle duration. The constituent steady-state probabilities, $p$ and $\tilde{\alpha}$ are given by \cite{10750858}:
\begin{equation}
    p = e^{\mathbb{W}_0\left(-\frac{n\lambda_{CB}}{1-n\lambda_{CB}(\tau_T-1)}\right)},
\end{equation}
\begin{equation}
    \tilde{\alpha} = \frac{1}{1-\lambda_{CB}(\tau_T-1)} \cdot \frac{1}{1-(\tau_T-1)p\ln p}.
\end{equation}

Our objective is to find the optimal packet size $L$ that minimizes $\overline{T}$ for each scheme. This optimization is subject to the fundamental constraint that the queue must remain unsaturated. The minimum required packet size, $L_{\min}$, and the corresponding feasibility conditions on network parameters (i.e., the constraints that network parameters must satisfy to ensure the network remains unsaturated) for each scheme are derived by applying the unsaturation condition (44) from \cite{10750858} to our specific models. The detailed derivation of these constraints is omitted for brevity. The resulting optimization problems are formulated as follows:

\textbf{Problem 1 (Connection-free Scheme):}
{\small
\begin{equation}\label{cfProblem}
\begin{aligned}
\min_{L} \quad & \overline{T}_{CF} = \left( \frac{\lambda_{CF} \overline{D}_{CF}^2 - \lambda_{CF} \overline{D}_{CF}}{2(1 - \lambda_{CF} \overline{D}_{CF})} + \overline{D}_{CF} \right) \cdot \left( \frac{L}{R} + \Delta_{CF} \right) \\
\text{s.t.} \quad & L \geq L_{\min, CF} = \left\lceil \frac{\Delta_{CF} \lambda_b R e^{nq}}{qR - \lambda_b e^{nq}} \right\rceil.
\end{aligned}
\end{equation}
}

(\ref{cfProblem}) is feasible only if the network parameters satisfy the condition $qR > \lambda_b e^{nq}$, which arises from the analysis ensuring the network remains unsaturated.

\textbf{Problem 2 (Connection-based Scheme):}
{\small
\begin{equation}\label{cbProblem}
\begin{aligned}
\min_{L} \quad & \overline{T}_{CB} = \left( \frac{\lambda_{CB} \overline{D}_{CB}^2 - \lambda_{CB} \overline{D}_{CB}}{2(1 - \lambda_{CB} \overline{D}_{CB})} + \overline{D}_{CB} \right) \cdot \Delta_{CB}^{F} \\
\text{s.t.} \quad & L \geq L_{\min, CB} = \left\lceil \frac{R(\lambda_b \Delta_{CB}^{F} e^{nq} + nq\lambda_b(\Delta_{CB}^{S}-\Delta_{CB}^{F}))}{q(R-n\lambda_b)} \right\rceil.
\end{aligned}
\end{equation}
}

Similarly, (\ref{cbProblem})  is feasible only if the network parameters satisfy the condition $R > n\lambda_b$.

\section{Optimal Packetization}\label{sectionIII}
In this section, we first employ numerical methods to identify the optimal packet size $L^*$ to achieve the minimum mean queueing delay (measured in seconds) $\overline T^*$,  for both connection-free and connection-based scheme. Next, we investigate the impact of various network parameters on $L^*$ and $\overline T^*$, and compare the sensitivity of the connection-free and connection-based schemes to these parameters.  Finally, we further explore the relationship between jitter of queueing and packet size $L$ through simulation, and compare the optimal points for mean queueing delay and jitter of queueing delay.

\subsection{Numerical Analysis of Optimal Packet Size}

\begin{figure}[htbp]
    \centering
    \subfigure[$\overline{T}$ versus $L$ for connection-free scheme.]{
        \label{meanCF}
        \begin{minipage}{0.9\columnwidth}
            \centering
            \includegraphics[width=\linewidth]{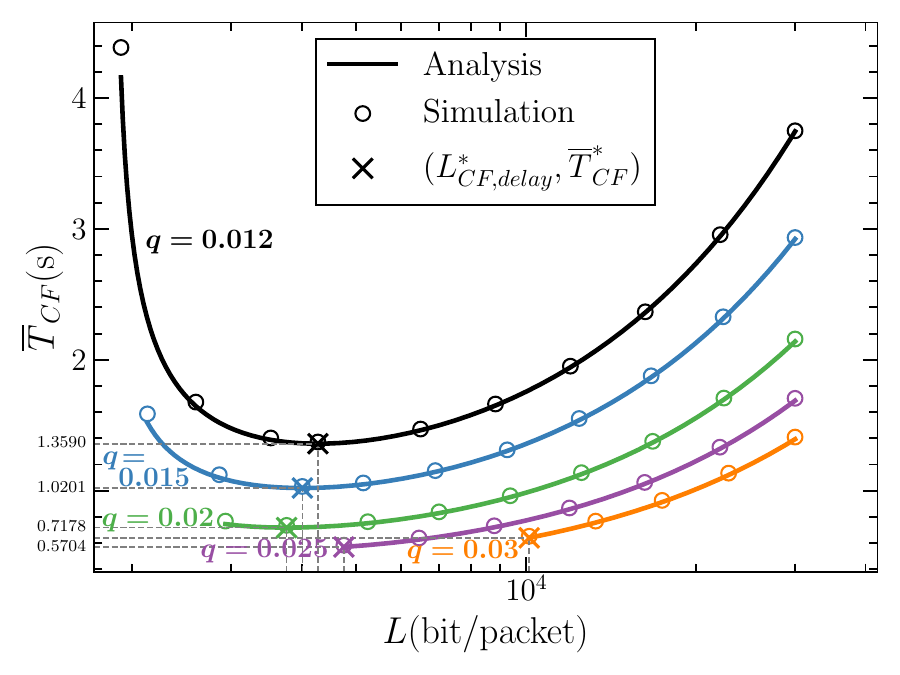}
        \end{minipage}
    }
    \vspace{1.5ex}
    \subfigure[$\overline{T}$ versus $L$ for connection-based scheme.]{
        \label{meanCB}
        \begin{minipage}{0.9\columnwidth}
            \centering
            \includegraphics[width=\linewidth]{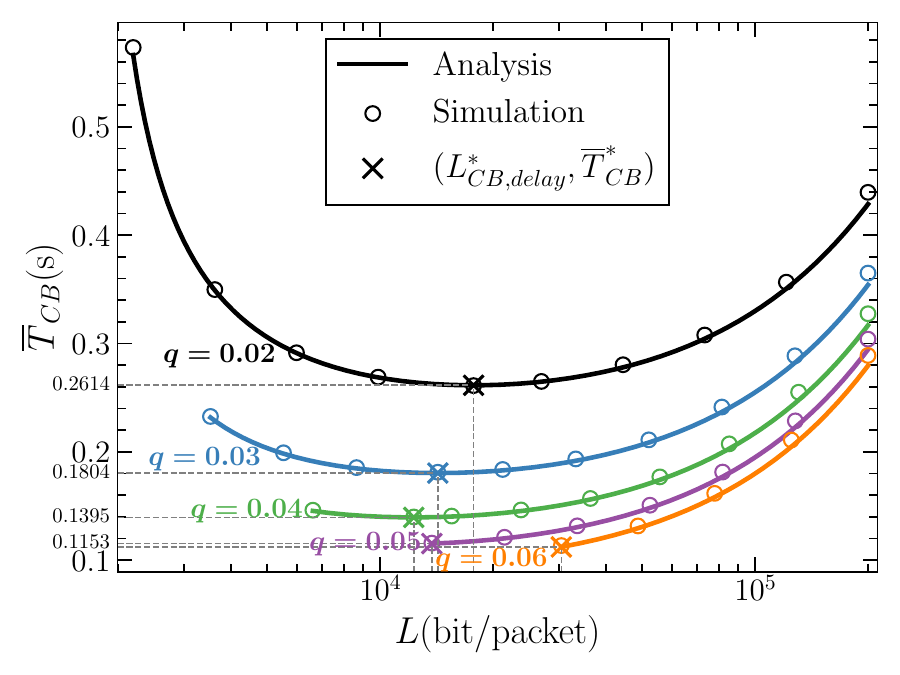}
        \end{minipage}
    }
    \caption{Mean queueing delay $\overline{T}$ versus packet size $L$ for (a) the connection-free scheme and (b) the connection-based scheme under different transmission probabilities $q$. The simulation results are obtained over a duration of $5 \times 10^4$ s. Common parameters are set as $n=100$, $\lambda_b = 10^3$ bit/s, and $R = 10^6$ bit/s. For the connection-free scheme, $\Delta_{CF}=0.005$ s. For the connection-based scheme, $\Delta_{CB}^{F}=0.004$ s and $\Delta_{CB}^{S}=0.009$ s.}
    \label{meanCFAndCB}
\end{figure}

\begin{figure*}[htbp] 
    \centering 
    \subfigure[$\overline T^*$ and $L^*$ versus $n$.]{
        \label{comp n}
        \begin{minipage}{0.3\textwidth}
            \centering
            \includegraphics[width=\linewidth]{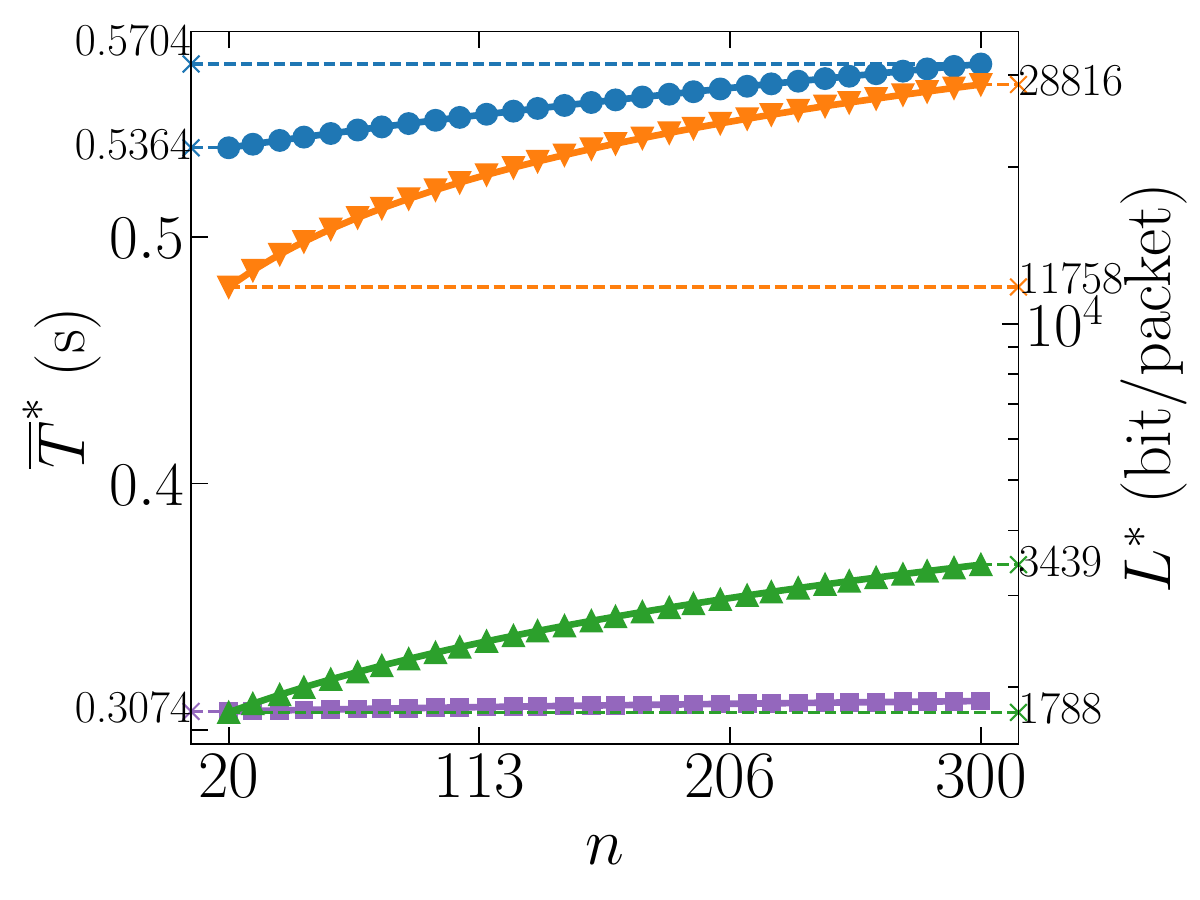}
        \end{minipage}
    }
    \hfill
    \subfigure[$\overline T^*$ and  $L^*$ versus $\lambda_b$.]{
        \label{comp lambda_b}
        \begin{minipage}{0.3\textwidth}
            \centering
            \includegraphics[width=\linewidth]{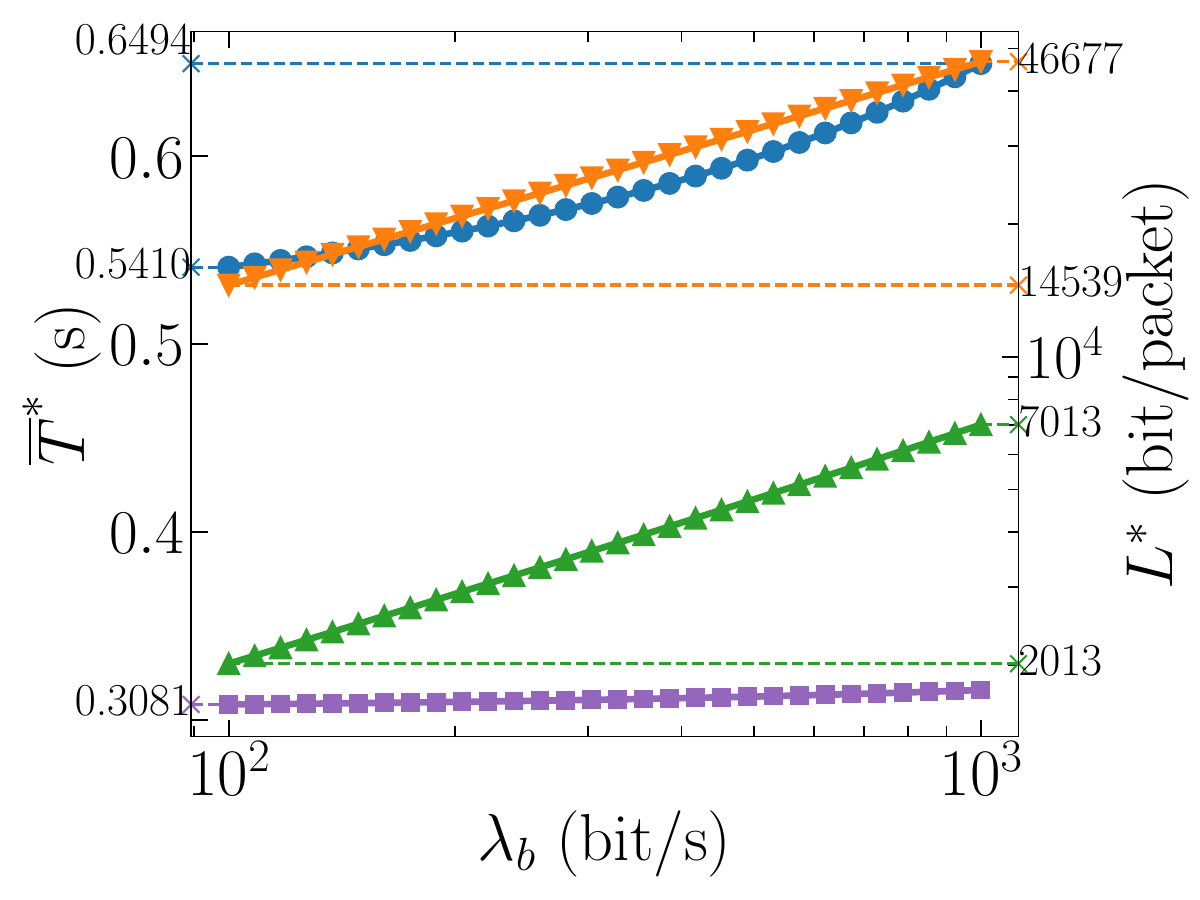}
        \end{minipage}
    }
    \hfill
    \subfigure[$\overline T^*$ and  $L^*$ versus $R$.]{
        \label{comp R}
        \begin{minipage}{0.3\textwidth}
            \centering
            \includegraphics[width=\linewidth]{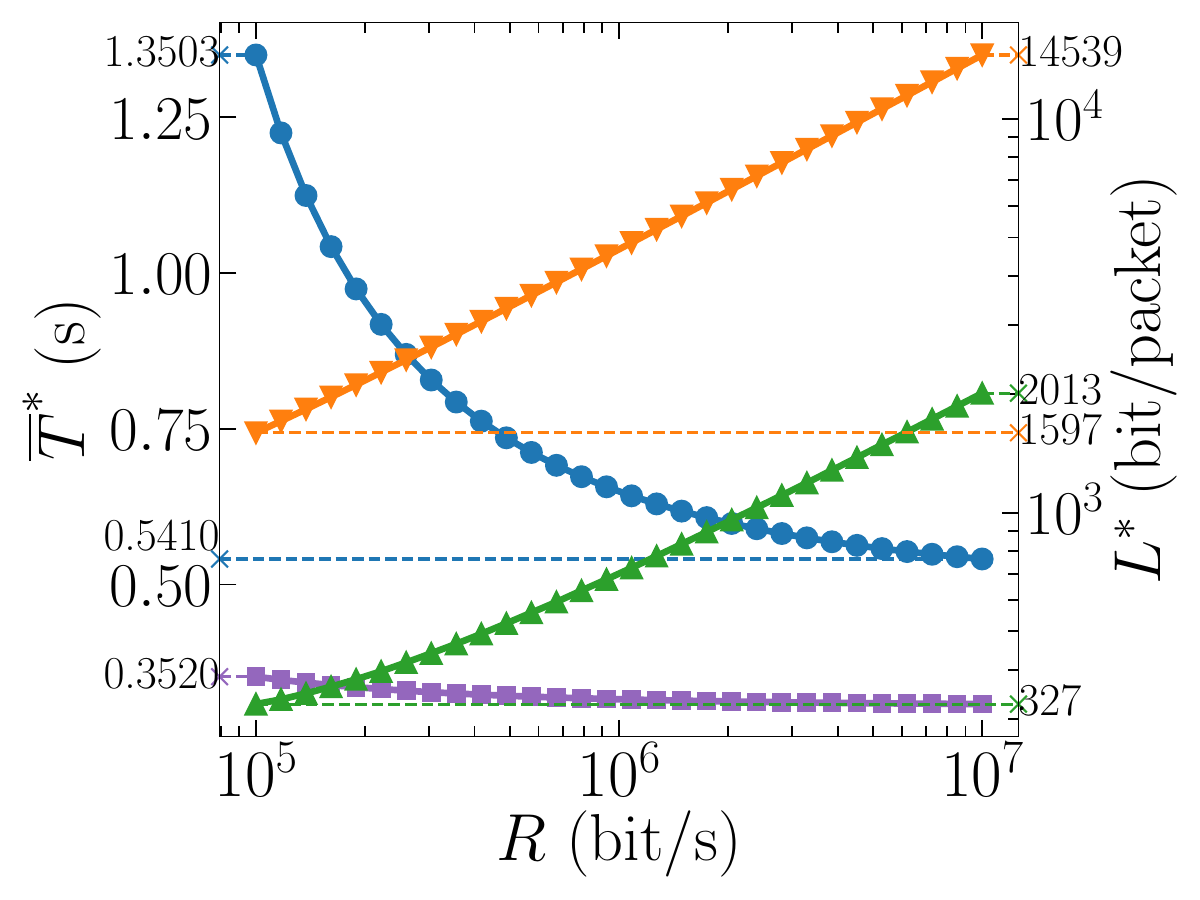}
        \end{minipage}
    }

    \vspace{1.5ex}

    \subfigure[$\overline T^*$ and $L^*$ versus the ACK duration for confirming data packet.]{
        \label{comp delta}
        \begin{minipage}{0.3\textwidth}
            \centering
            \includegraphics[width=\linewidth]{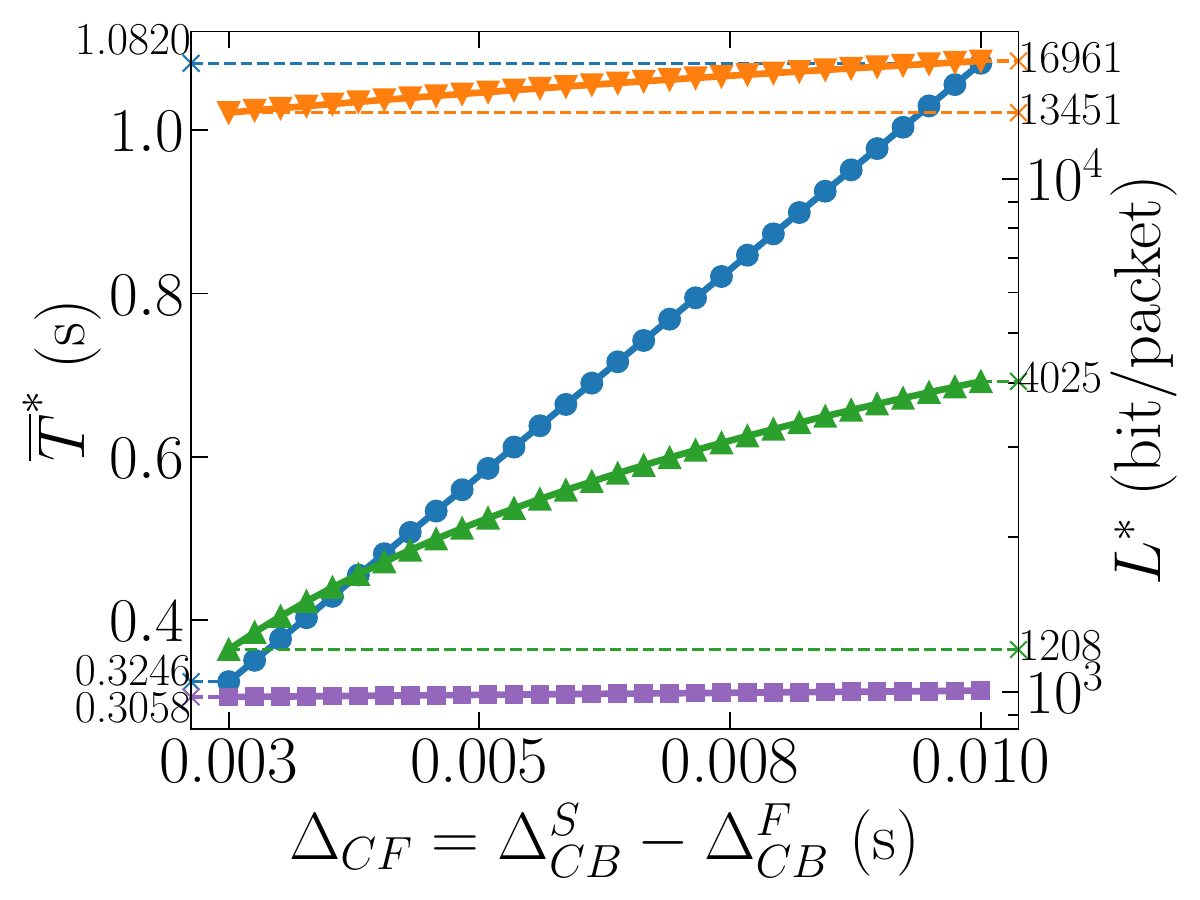}
        \end{minipage}
    }
    \hfill
    \subfigure[$\overline T^*$ and $L^*$ versus $\Delta_{CB}^{F}$.]{
        \label{comp deltaF}
        \begin{minipage}{0.3\textwidth}
            \centering
            \includegraphics[width=\linewidth]{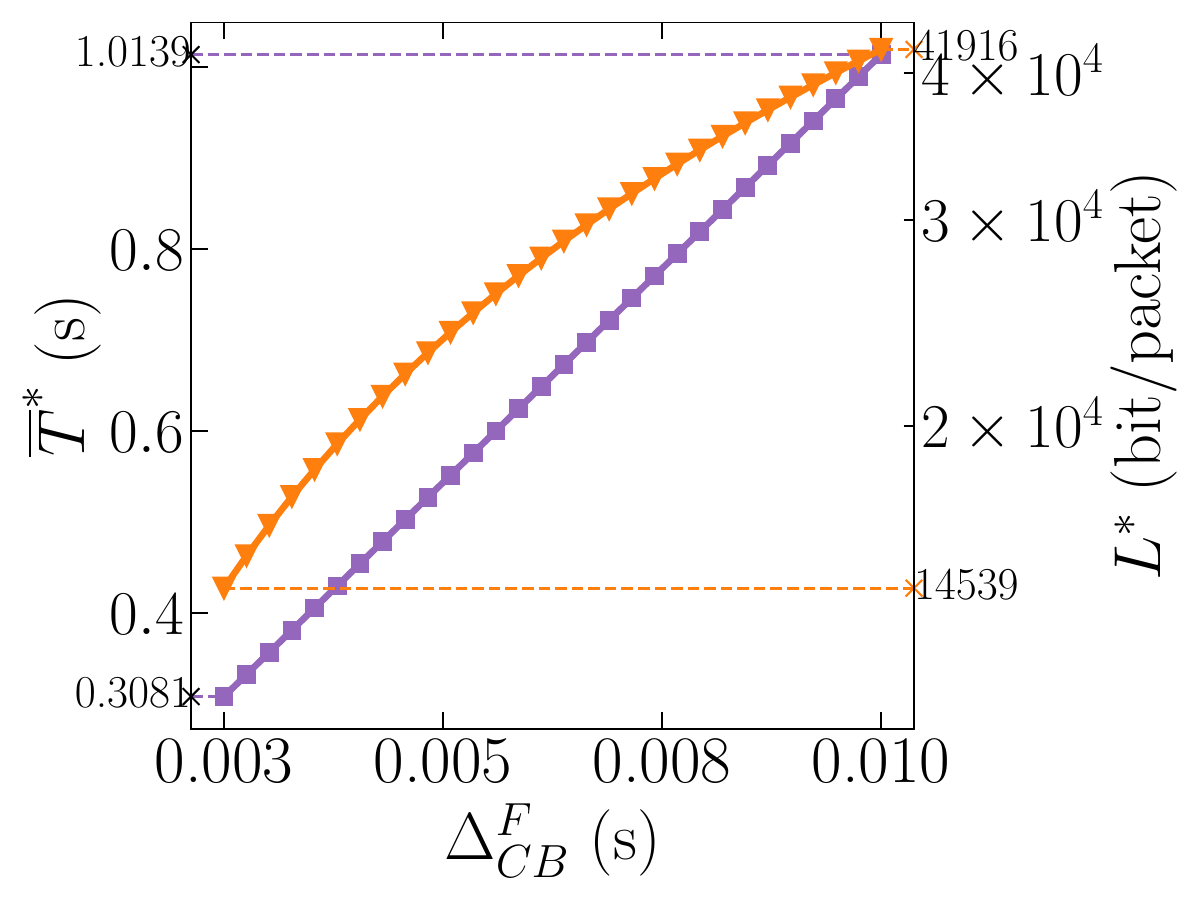}
        \end{minipage}
    }
    \hfill
    \begin{minipage}{0.3\textwidth}
            \centering
            \includegraphics[width=\linewidth]{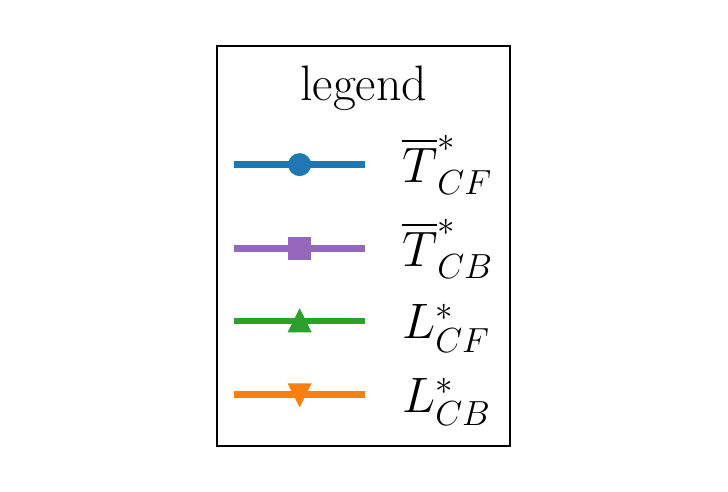}
    \end{minipage}
    \caption{The optimal mean queueing delay $\overline T^*$ and the corresponding packet size $L^*$ versus various network parameters. The default parameters are set as: $n=50$, $\lambda_b=10^2$ bit/s, $q=0.01$, $R=10^7$ bit/s, $\Delta_{CF}=0.005$ s, $\Delta_{CB}^{F}=0.003$ s, and $\Delta_{CB}^{S}=0.008$ s.}
    \label{comp}

\end{figure*} 

The mean queueing delays (measured in seconds) 
 $\overline{T}_{{CF}}$ and $\overline{T}_{{CB}}$ are highly complicated functions of the packet size $L$. Although under certain conditions, such as for the connection-free scheme when $n\lambda$ is small, an approximate explicit expression for the optimal packet size $L^*$ can be derived using the approximation method involving $e^{\mathbb{W}_0(-n\lambda_{{CF}})}$ provided in Equation (75) of \cite{6205590}, obtaining an explicit expression for $L^*$ is generally challenging. Consequently, we uniformly adopt numerical methods (specifically, an exhaustive search over the integer space of $L$) to analyze $L^*$.

Fig.~\ref{meanCFAndCB} illustrates the variation of $\overline{T}_{{CF}}$ and $\overline{T}_{{CB}}$ with respect to $L$ under different values of $q$. As shown in Fig.~\ref{meanCFAndCB}, both connection-free and connection-based schemes exhibit similar trends. 
For relatively smaller values of $q$ ($q=0.012$, $0.015$, $0.02$ for the connection-free scheme, and $q=0.02$, $0.03$, $0.04$ for the connection-based scheme), within the range of $L$ that ensures the network remains unsaturated, the mean queueing delay first decreases monotonically and then increases monotonically with increasing $L$, with the extremum point corresponding to the optimal packet size $L^*$. This non-monotonic trend clearly validates the  relationship between packetization and queueing delay as described in Section~\ref{SectionI}. 
For relatively larger values of $q$ ($q=0.025$, $0.03$ for the connection-free scheme, and $q=0.05$, $0.06$ for the connection-based scheme), within the range of $L$ that ensures the network remains unsaturated, the mean queueing delay increases monotonically with increasing $L$, and in this case, $L^*$ equals the minimum packet size $L_{\min}$. This behavior occurs because the optimal packet size $L^*$ is determined by the interplay of two factors: the theoretical delay extremum $L_{0}$ and $L_{\min}$. The actual optimal choice is therefore given by $L^* = \max(L_{\min}, L_{0})$. As $q$ increases, $L_{0}$ gradually decreases while $L_{\min}$ (which itself has a non-monotonic relationship with $q$) eventually surpasses $L_{0}$. Another key observation from Fig.~\ref{meanCFAndCB} is that for valid $L$ values, $\overline{T}$ consistently decreases with increasing $q$ in both schemes. Additionally, the simulation results in Fig.~\ref{meanCFAndCB} closely match the theoretical curves, validating the accuracy of the theoretical analysis.
\subsection{Parametric Sensitivity Analysis}

Fig.~\ref{comp} demonstrates how the optimal mean queueing delay $\overline{T}^*$ and the corresponding optimal packet size $L^*$ vary with different network parameters, including the number of nodes $n$, the bit arrival rate $\lambda_b$, the uplink transmission rate $R$, the ACK duration for confirming data packets, and the request duration in connection-based scheme $\Delta_{{CB}}^F$.
It is noteworthy that for the specific parameter settings of Fig.~\ref{comp}, the connection-based scheme consistently yields lower optimal queueing delays than its connection-free counterpart. However, as will be demonstrated in the following section, this observation does not hold universally. The primary objective of the current analysis is to understand the variation trends of both the optimal delay and its corresponding optimal packet size with respect to different network parameters, as well as to examine the sensitivity differences between connection-based and connection-free schemes under parameter variations. A comprehensive trade-off analysis comparing these two schemes will be presented in detail in the next section.

As shown in Fig.~\ref{comp n}, as $n$ increases, $\overline{T}^{*}_{CF}$, $\overline{T}^{*}_{CB}$, $L^*_{CF}$, and $L^*_{CB}$ all monotonically increase. Notably, the growth rate of $\overline{T}^{*}_{CF}$ is significantly larger than that of $\overline{T}^{*}_{CB}$. When $n$ increases from 20 to 300, $\overline{T}^{*}_{CF}$ rises from 0.5364 s to 0.5704 s, whereas $\overline{T}^{*}_{CB}$ shows only a marginal increase from its initial value of 0.3074 s. In contrast to the optimal delay, the growth rate of $L^*_{CB}$ is larger than that of $L^*_{CF}$ (note that the right-hand $L^*$-axis is on a logarithmic scale). $L^*_{CB}$ grows from 11758 bit/packet to 28816 bit/packet, representing a 145.1\% increase.  However, $L^*_{CF}$ only increases from 1788 bit/packet to 3439 bit/packet, a growth of 92.3\%. In summary, as $n$ increases, the connection-based scheme demonstrates stronger robustness in delay performance while exhibiting a more rapid increase in $L$ compared to the connection-free scheme.  The impact of $\lambda_b$ shown in Fig.~\ref{comp lambda_b} is similar to that of $n$.

As shown in Fig.~\ref{comp R}, $\overline{T}^{*}$ and $L^{*}$ exhibit opposite trends with increasing $R$. While $\overline{T}^{*}_{CF}$ and $\overline{T}^{*}_{CB}$ decrease monotonically, interestingly, $L^{*}_{CF}$ and $L^{*}_{CB}$ continue to increase monotonically, following the same pattern observed for variations in $n$ and $\lambda_b$. The reduction in $\overline{T}^{*}_{CF}$ is significantly greater than that in $\overline{T}^{*}_{CB}$, indicating that the connection-free scheme benefits more from increased $R$ compared to the connection-based scheme. On the other hand, $L^{*}_{CB}$ still shows a larger increase than $L^{*}_{CF}$.

Fig.~\ref{comp delta} demonstrates the impact of the ACK duration for confirming data packets (i.e.,  $\Delta_{CF}$ for the connection-free scheme and $(\Delta_{CB}^{S}-\Delta_{CB}^{F})$ for the connection-based scheme). As observed from Fig.~\ref{comp delta}, although both $\overline{T}^{*}$ and $L^{*}$ increase monotonically, a consistent and significant difference between the two schemes is observed in their sensitivity to this duration, a finding that holds true for both $\overline{T}^{*}$ and $L^{*}$. The connection-free scheme shows high sensitivity to this duration, while the connection-based scheme remains largely insensitive. When the duration increases from 0.003 s to 0.01 s, $\overline{T}^{*}_{CB}$ remains stable at approximately 0.3058 s, and $L^{*}_{CB}$ increases from 13451 bit/packet to 16961 bit/packet (a 26.1\% increase). In contrast, $\overline{T}^{*}_{CF}$ increases substantially from 0.3246 s to 1.0820 s (a 233.3\% increase), and $L^{*}_{CF}$ increases from 1208 bit/packet to 4025 bit/packet (a 233.2\% increase). Fig.~\ref{comp deltaF} illustrates the impact of $\Delta_{CB}^{F}$ on the connection-based scheme. Unlike the connection-based scheme's performance in Fig.~\ref{comp delta}, it now shows strong sensitivity. With the same horizontal axis range (0.003 s to 0.01 s), $\overline{T}^{*}_{CB}$ increases significantly from 0.3081 s to 1.0139 s, and $L^{*}_{CB}$ also increases substantially from 14539 bit/packet to 41916 bit/packet. The above phenomenon occurs because, for the connection-free scheme, the ACK duration for confirming data packets appears in every transmission attempt (regardless of success or failure), whereas for the connection-based scheme, this duration only occurs during successful transmissions when data packets are formally delivered. On the other hand, for the connection-based scheme, $\Delta_{CB}^{F}$ appears in every request attempt.

\subsection{Jitter of Queueing Delay}

\begin{figure}[htbp]
    \centering
    \subfigure[$J$ versus $L$ for the connection-free scheme.]{
        \label{jitterCF}
        \begin{minipage}{\columnwidth}
            \centering
            \includegraphics[width=0.9\linewidth]{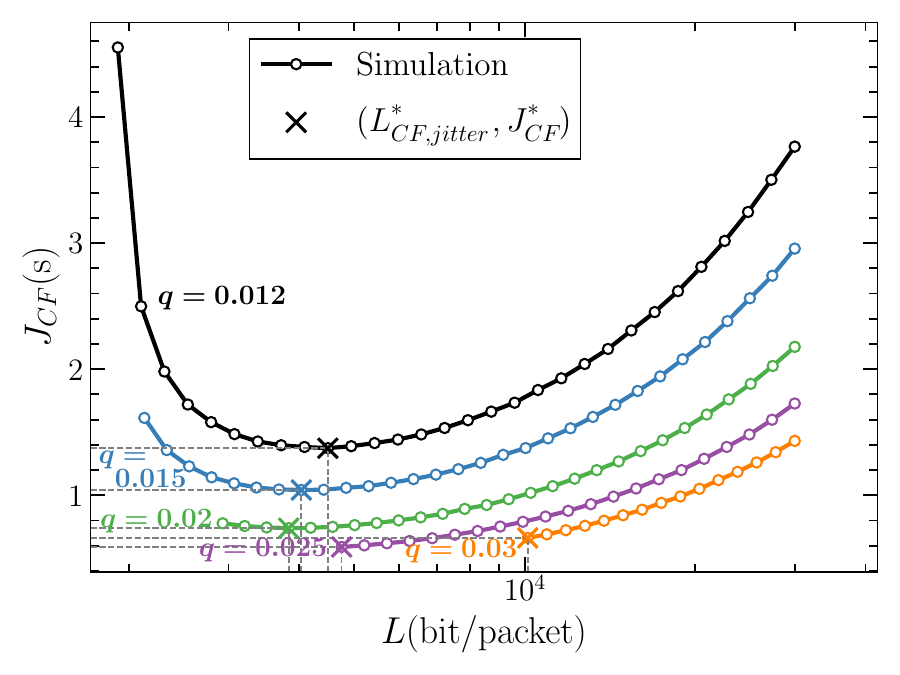}
        \end{minipage}
    }
    \vspace{1.5ex} 
    \subfigure[$J$ versus $L$ for the connection-based scheme.]{
        \label{jitterCB}
        \begin{minipage}{\columnwidth}
            \centering
            \includegraphics[width=0.9\linewidth]{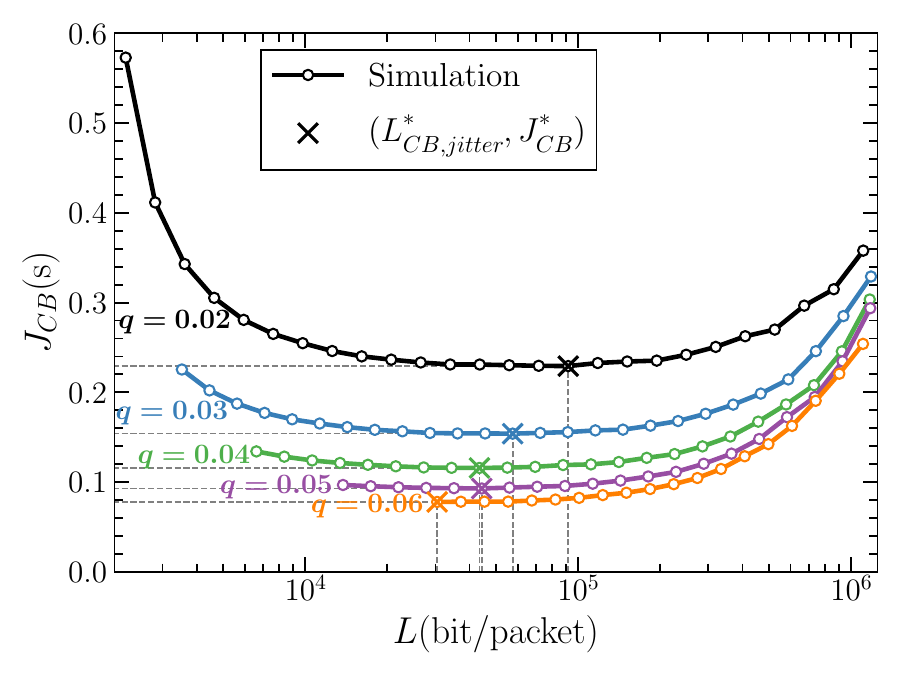}
        \end{minipage}
    }
    \caption{ Jitter of queueing delay $J$ versus packet size $L$ for (a) the connection-free scheme and (b) the connection-based scheme under different transmission probabilities $q$. Common parameters are set as $n=100$, $\lambda_b = 10^3$ bit/s, and $R = 10^6$ bit/s. For the connection-free scheme (a), $\Delta_{CF}=0.005$ s and the simulation is run over a duration of $5 \times 10^4$ s. For the connection-based scheme (b), $\Delta_{CB}^{F}=0.004$ s, $\Delta_{CB}^{S}=0.009$ s, and the simulation is run over a duration of $2 \times 10^5$ s.}
    \label{jitterSim}
\end{figure}

Since the queueing delay of each packet is random, in addition to the mean queueing delay, the variability of queueing delay (i.e., jitter) represents another important performance metric, particularly for audio/video streaming applications. Smaller jitter indicates a more stable network and enables better prediction of each packet's delay. Therefore, following \cite{9244230}, we similarly further examine jitter of queueing delay, which is defined as the standard deviation of queueing delay measured in seconds, and denoted by $J$ in this study. In our model, since the service time and waiting time are interdependent with network parameters rather than being independent, it is challenging to derive a closed-form expression for jitter. We consequently adopt simulation approaches for investigation. Fig.~\ref{jitterSim} demonstrates the simulated variations of $J_{{CF}}$ and $J_{{CB}}$ with respect to $L$ across different $q$ values, using the same network parameter settings as in Fig.~\ref{meanCFAndCB}. As evident from Fig.~\ref{jitterSim}, jitter exhibits a similar relationship with packetization as observed between mean queueing delay and packetization, suggesting that jitter performance can likewise be optimized through careful packetization adjustment.

A comparison between Fig.~\ref{meanCF} and Fig.~\ref{jitterCF} reveals nearly identical variation trends between $\overline{T}_{CF}$ and $J_{CF}$. Furthermore, the comparison between $L^{*, {simulation}}_{CF, {jitter}}$ and $L^{*}_{CF, {delay}}$ shows that they are almost equal at $q$ =  0.012, 0.015, and 0.02, while both equal $L_{\text{min}}$ at $q$ = 0.025 and 0.03. This demonstrates that for the connection-free scheme, optimizing mean queueing delay and jitter of queueing delay are highly synergistic, allowing a single packetization strategy to effectively co-optimize both metrics.

In contrast, the connection-based scheme exhibits a more complex relationship, as evident from comparing Fig.~\ref{jitterCB} and Fig.~\ref{meanCB}. At $q$ = 0.06, both $L^{*, {simulation}}_{CB, {jitter}}$ and $L^{*}_{CB, {delay}}$ equal $L_{\text{min}}$. However, at $q$ = 0.02, 0.03, 0.04, and 0.05, $L^{*, {simulation}}_{CB, {jitter}}$ is significantly larger than $L^{*}_{CB, {delay}}$. For instance, at $q$ = 0.03, $L^{*}_{CB, {delay}}$ is approximately $1.5\times 10^4$ bit/packet, while $L^{*, {simulation}}_{CB, {jitter}}$ approaches $6\times 10^4$ bit/packet. Nevertheless, $J_{CB}$ exhibits a remarkably flat trough near $L^{*, {simulation}}_{CB, {jitter}}$, meaning the $J_{CB}$ at $L^{*}_{CB, {delay}}$ does not differ substantially from $J_{CB}^{*, {simulation}}$. Consequently, for the connection-based scheme,  while less perfectly aligned than in the connection-free case, optimizing mean queueing delay and jitter still maintains considerable synergy, enabling a packetization strategy optimized for mean queueing delay to achieve excellent jitter performance without significant compromise.

\section{Connection-free versus Connection-based Schemes}\label{sectionIV}

In this section, we first characterize the trade-off between connection-free and connection-based schemes based on the ratio of the ACK duration of the connection-free scheme to the request duration of the connection-based scheme, while accounting for the differences in optimal packet sizes between the two schemes. This analysis identifies three key threshold ratios and the four distinct operational regions they define, providing a criterion for selecting between connection-free and connection-based schemes. Subsequently, we investigate the impact of various network parameters on these three threshold ratios and the corresponding four regions.

\subsection{Trade-off between Connection-Free and Connection-Based Schemes}
Through a comprehensive comparison of Fig.~\ref{comp n} to Fig.~\ref{comp delta}, we conclude that $\Delta_{CF}$ is the most significant factor affecting the performance of the connection-free scheme, while $\Delta_{CB}^{F}$ is the most significant factor affecting the performance of the connection-based scheme. We further find that the ratio $\Delta_{CF}$/$\Delta_{CB}^{F}$ significantly influences the trade-off between connection-free and connection-based schemes. Additionally, as shown in Fig.~\ref{comp}, within the wide range of network parameters considered, the optimal packet size for the connection-based scheme is consistently larger than that for the connection-free scheme under the same network conditions. Therefore, we cannot simply compare their overall minimum delay, as this would overlook the significant difference in their optimal packet sizes and  lead to a loss of some valuable insights.

\begin{figure}[htbp]
    \centering    \includegraphics[width=\columnwidth]{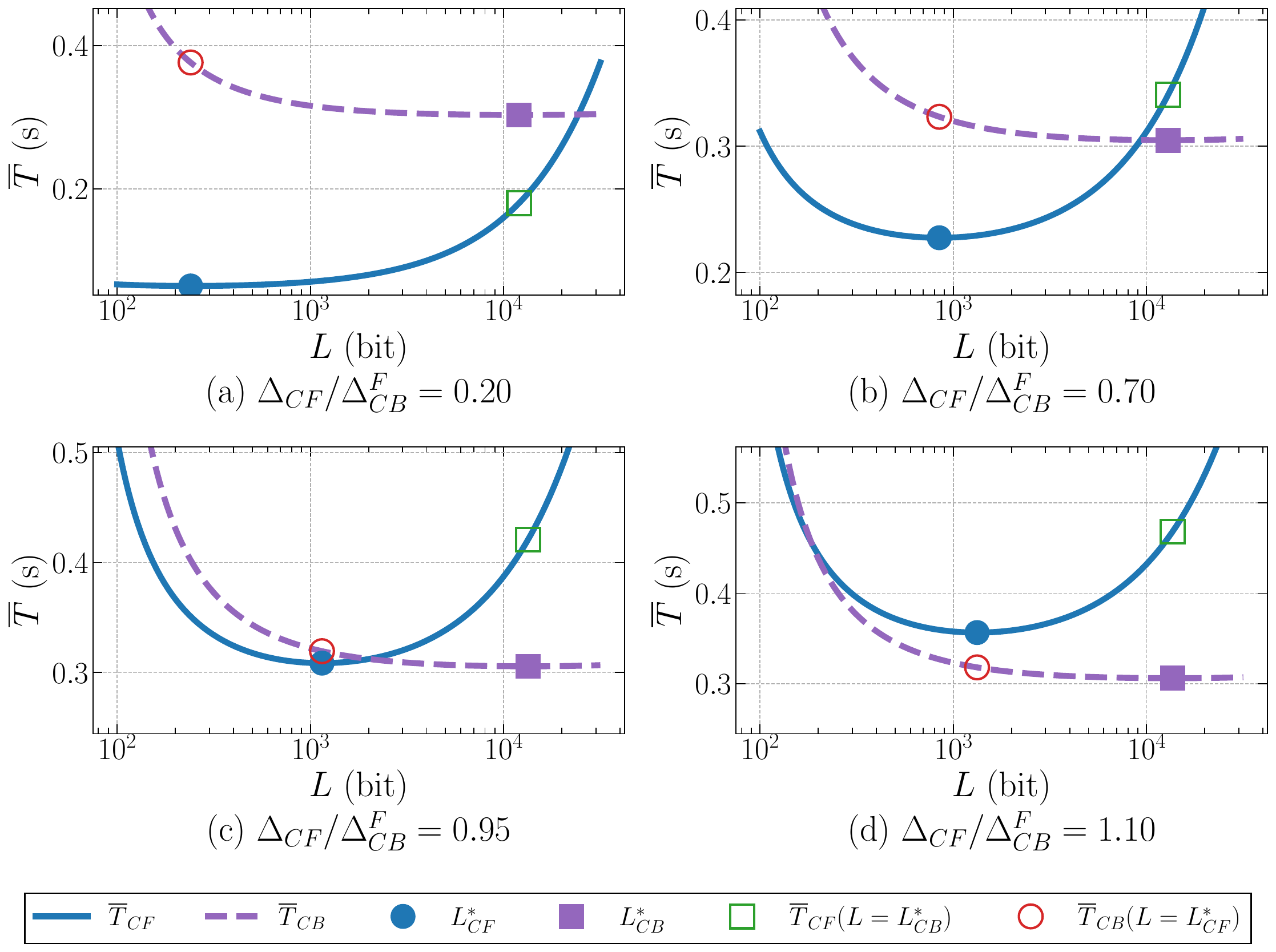}
    \caption{Mean queueing delay comparison of the connection-free scheme $\overline{T}_{CF}$ and connection-based scheme $\overline{T}_{CB}$ versus $L$ under several different $\Delta_{CF}/\Delta_{CB}^F$. $n=50$, $\lambda_b = 10^2$ bit/s, and $R = 10^7$ bit/s, $q=0.01$, $\Delta_{CB}^{F}=0.003$ s.}
    \label{preExp}
\end{figure}

Fig.~\ref{preExp} shows how the mean queueing delay of connection-based and connection-free schemes changes with $L$ under several different $\Delta_{CF}/\Delta_{CB}^{F}$ (where $\Delta_{CB}^{F}$ is held constant while $\Delta_{CF}$ varies across different values, and for the considered $\Delta_{CF}$/$\Delta_{CB}^{F}$, $L^*_{CB}>L^*_{CF}$ holds). From Fig.~\ref{preExp}(a), we can see that when the $\Delta_{CF}/\Delta_{CB}^{F}$ is small, the connection-free scheme achieves a lower overall minimum delay, i.e., $\overline T_{CF}^*<\overline T_{CB}^*$. Moreover, in this case, the advantage of the connection-free scheme is so pronounced that even near the optimal packet size of the connection-based scheme $L_{CB}^*$, the connection-free scheme still provides a lower delay, i.e., $\overline T_{CF}^{L=L_{CB}^*}<\overline T_{CB}^*$. Only at larger packet sizes does the delay of the connection-free scheme become greater than that of the connection-based scheme. As the $\Delta_{CF}/\Delta_{CB}^{F}$ increases, as shown in Fig.~\ref{preExp}(b), the connection-free scheme still achieves a lower overall minimum delay, but in this case, the advantage of the connection-free scheme is not as pronounced as in Fig.~\ref{preExp}(a), while the advantage of the connection-based scheme becomes more evident near the optimal packet size of the connection-based scheme $L_{CB}^*$, where its delay is now worse than that of the connection-based scheme, i.e., $\overline T_{CF}^{L=L_{CB}^*}>\overline T_{CB}^*$. As the $\Delta_{CF}/\Delta_{CB}^{F}$ increases further, shown in Fig.~\ref{preExp}(c), the advantage of the connection-free scheme diminishes further, and the connection-free scheme can no longer achieve a lower overall minimum delay, i.e., $\overline T_{CF}^{*} > \overline T_{CB}^*$. However, the connection-free scheme still retains some advantage because near the optimal packet size of the connection-free scheme $L_{CF}^*$, the delay of the connection-based scheme is still worse than that of the connection-free scheme, i.e., $\overline T_{CF}^{*}<\overline T_{CB}^{L=L_{CF}^*}$. When the $\Delta_{CF}/\Delta_{CB}^{F}$ is very large, as shown in Fig.~\ref{preExp}(d), at this point, the advantage of the connection-based scheme becomes so pronounced that not only does the connection-free scheme fail to achieve a lower overall minimum delay, but the connection-based scheme also performs better even near the optimal packet size of the connection-free scheme, i.e. $\overline T^*_{CB}<\overline T_{CF}^*$ and $\overline T_{CB}^{L=L_{CF}^*} < \overline T_{CF}^{*}$. Only for even smaller packet sizes does the connection-based scheme's delay become larger than the connection-free scheme's (it is worth noting that further experiments show this crossover at small $L$ may not always exist).

Drawing from the four different advantage cases for connection-based and connection-free schemes presented in Fig.~\ref{preExp}, we further present Fig.~\ref{fourRegions} to illustrate the trade-off between them. The $x$-axis of Fig.~\ref{fourRegions} represents the ratio $\Delta_{CF}$/$\Delta_{CB}^{F}$ (where $\Delta_{CB}^{F}$ is also held constant while $\Delta_{CF}$ varies across different values, and within the considered range of $\Delta_{CF}$/$\Delta_{CB}^{F}$, $L^*_{CB}>L^*_{CF}$ always holds), and the figure includes four curves depicting the variation with respect to this ratio:  $\overline{T}^*_{CF}$,  $\overline{T}^*_{CB}$, $\overline{T}_{CF}^{L=L^*_{CB}}$, and $\overline{T}_{CB}^{L=L^*_{CF}}$.

The three key threshold ratios in Fig.~\ref{fourRegions} are defined as follows: $\xi_1$ is the value of $\Delta_{CF}/\Delta_{CB}^{F}$ at which $\overline{T}_{CB}^* = \overline{T}_{CF}^{L=L_{CB}^*}$; $\xi_2$ is the value of $\Delta_{CF}/\Delta_{CB}^{F}$ at which $\overline{T}_{CF}^* = \overline{T}_{CB}^*$; and $\xi_3$ is the value of $\Delta_{CF}/\Delta_{CB}^{F}$ at which $\overline{T}_{CF}^* = \overline{T}_{CB}^{L=L_{CF}^*}$. These three threshold ratios divide the parameter space into four regions, defined as follows:

\begin{figure}[htbp]
    \centering
    \includegraphics[width=\columnwidth]{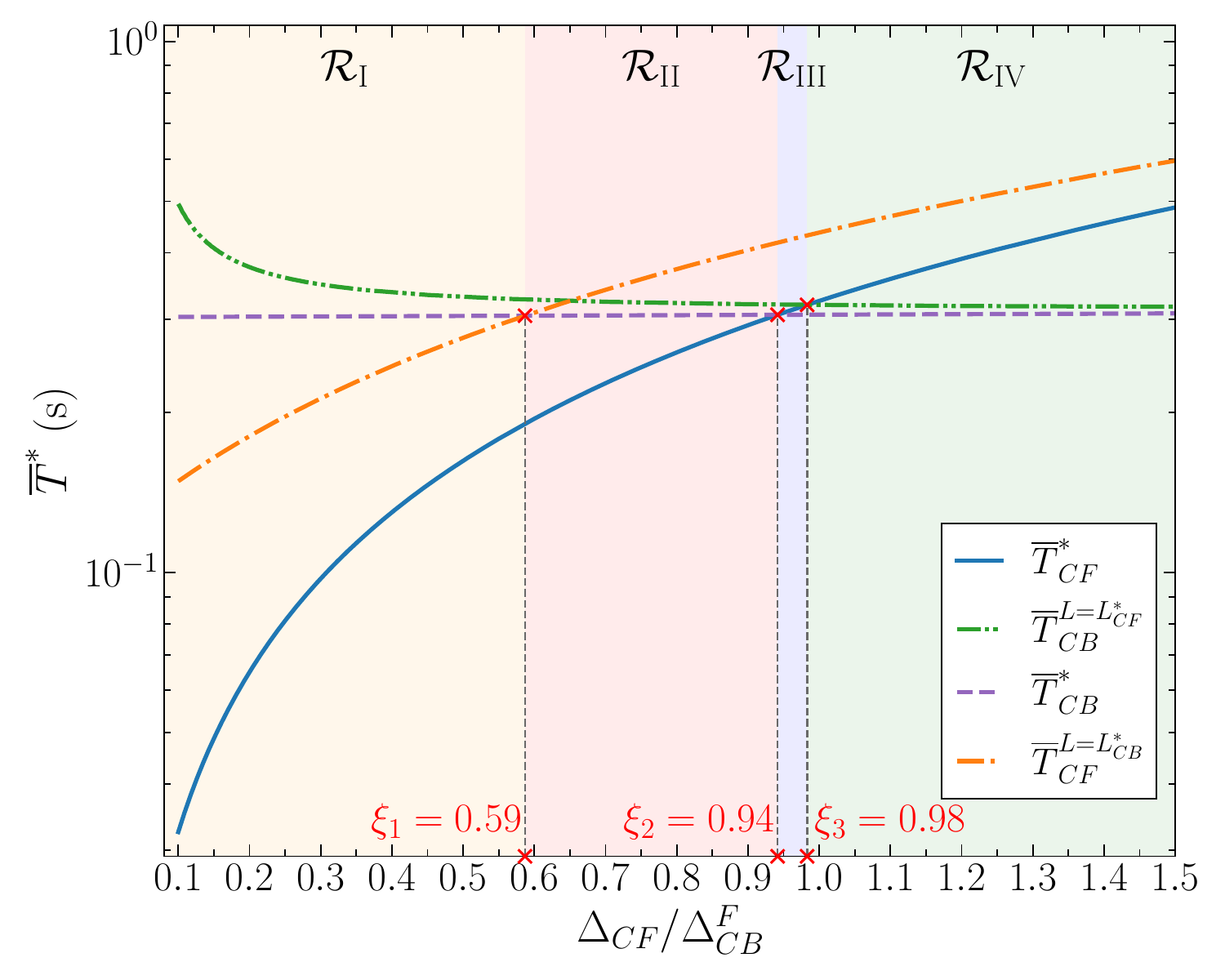}
    \caption{Performance trade-off between the connection-free and connection-based schemes. The parameter space of  $\Delta_{CF}/\Delta_{CB}^{F}$ is divided into four distinct regions $\mathcal{R}_\text {I}, \mathcal{R}_\text{II}, \mathcal{R}_\text{III}$ and $\mathcal{R}_\text{IV}$, bounded by three threshold ratios $\xi_1, \xi_2,$ and $\xi_3$. $n=50$, $\lambda_b = 10^2$ bit/s, and $R = 10^7$ bit/s, $q=0.01$, $\Delta_{CB}^{F}=0.003$ s.}
    \label{fourRegions}
\end{figure}

\begin{figure*}[htbp]
    \centering
    \subfigure[Impact of the number of nodes $n$.]{
        \label{regions n}
        \begin{minipage}{0.45\textwidth}
            \centering
            \includegraphics[width=\linewidth]{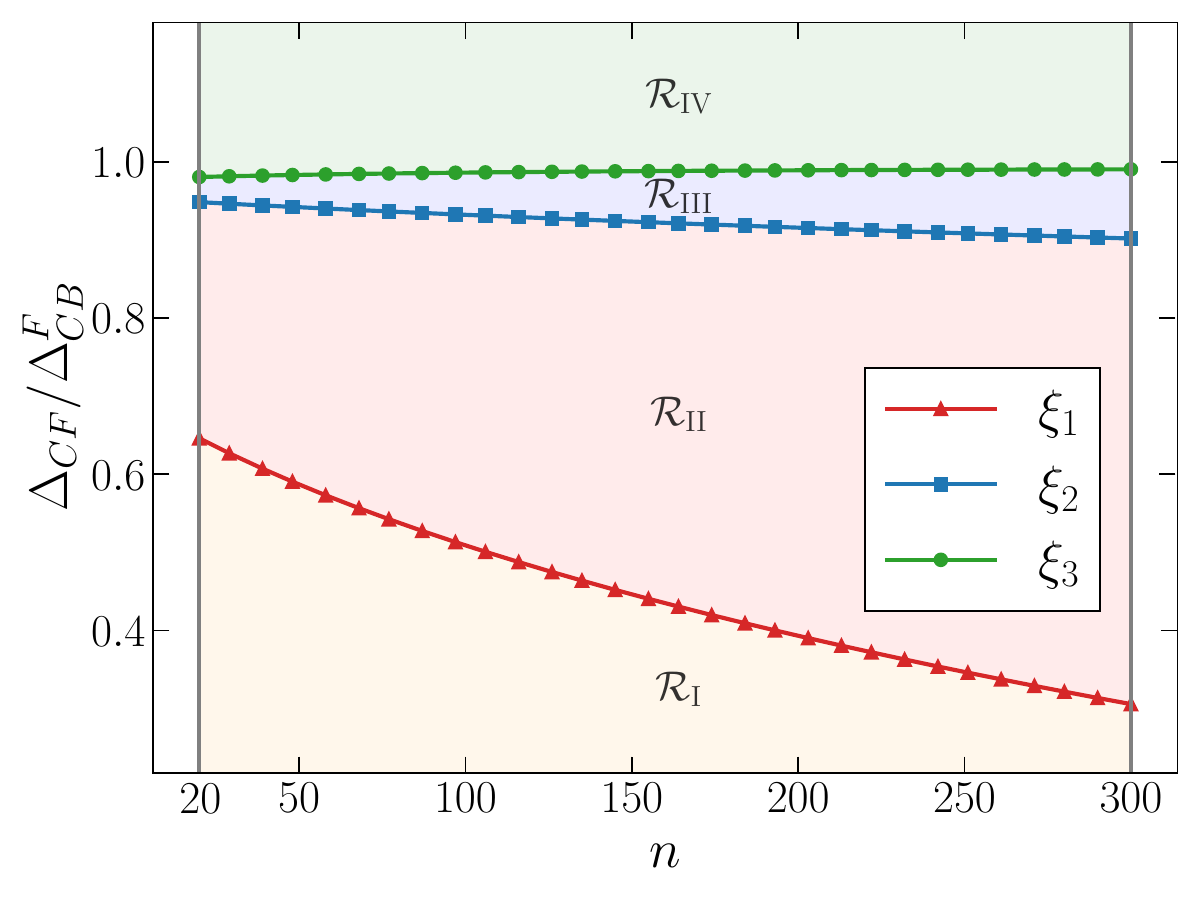}
        \end{minipage}
    }
    \hfill
    \subfigure[Impact of the bit arrival rate $\lambda_b$.]{
        \label{regions lambda_b}
        \begin{minipage}{0.45\textwidth}
            \centering
            \includegraphics[width=\linewidth]{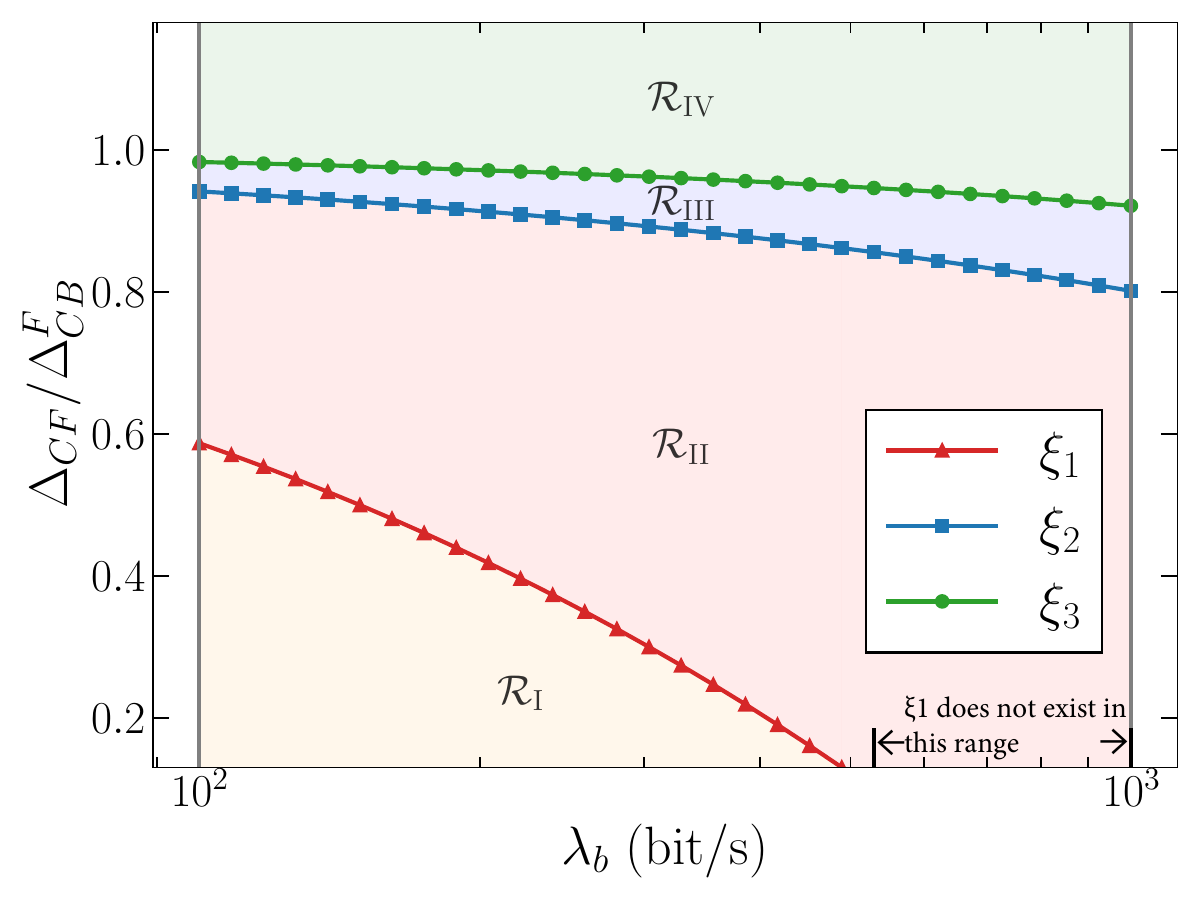}
        \end{minipage}
    }

    \vspace{1.5ex}

    \subfigure[Impact of the uplink data rate $R$.]{
        \label{regions R}
        \begin{minipage}{0.45\textwidth}
            \centering
            \includegraphics[width=\linewidth]{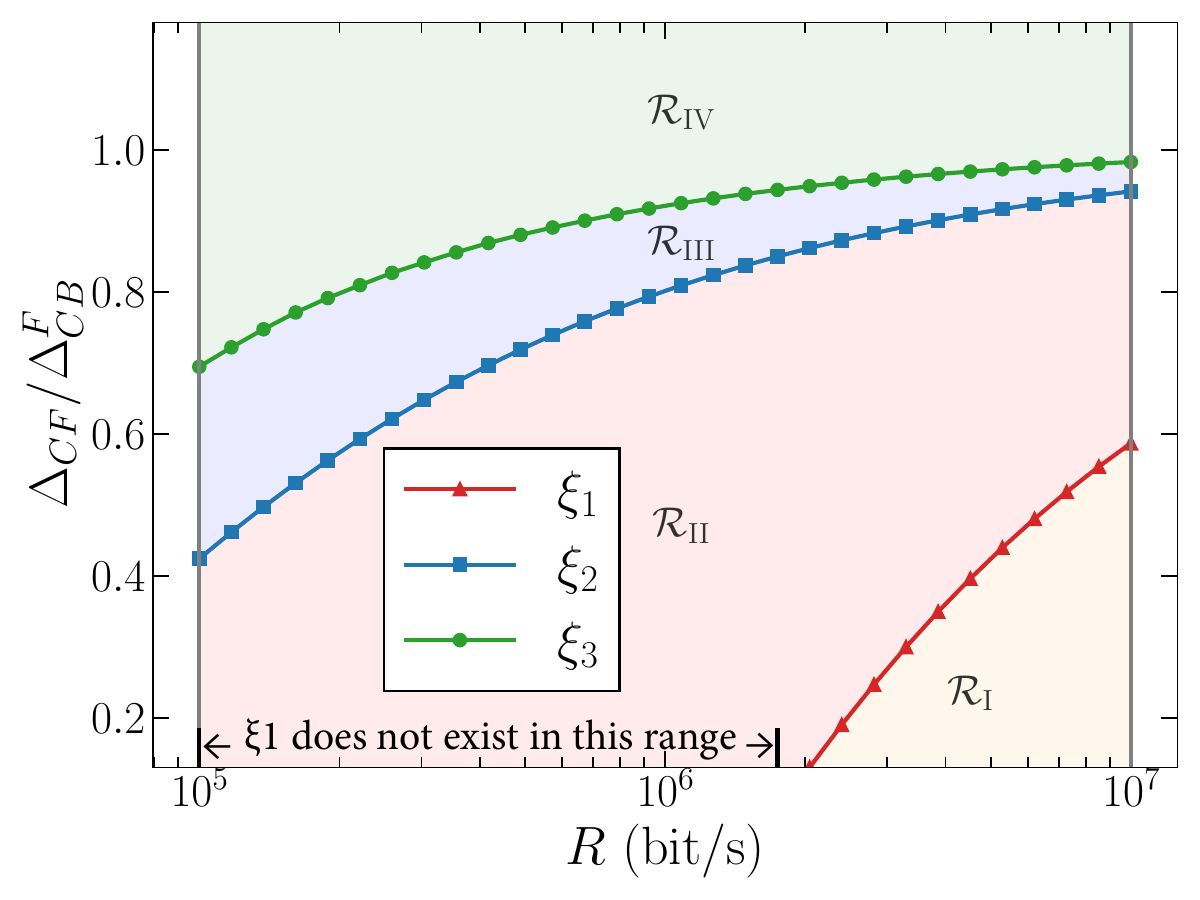}
        \end{minipage}
    }
    \hfill
    \subfigure[Impact of connection-based scheme request duration $\Delta_{CB}^{F}$.]{
        \label{regions deltaF}
        \begin{minipage}{0.45\textwidth}
            \centering
            \includegraphics[width=\linewidth]{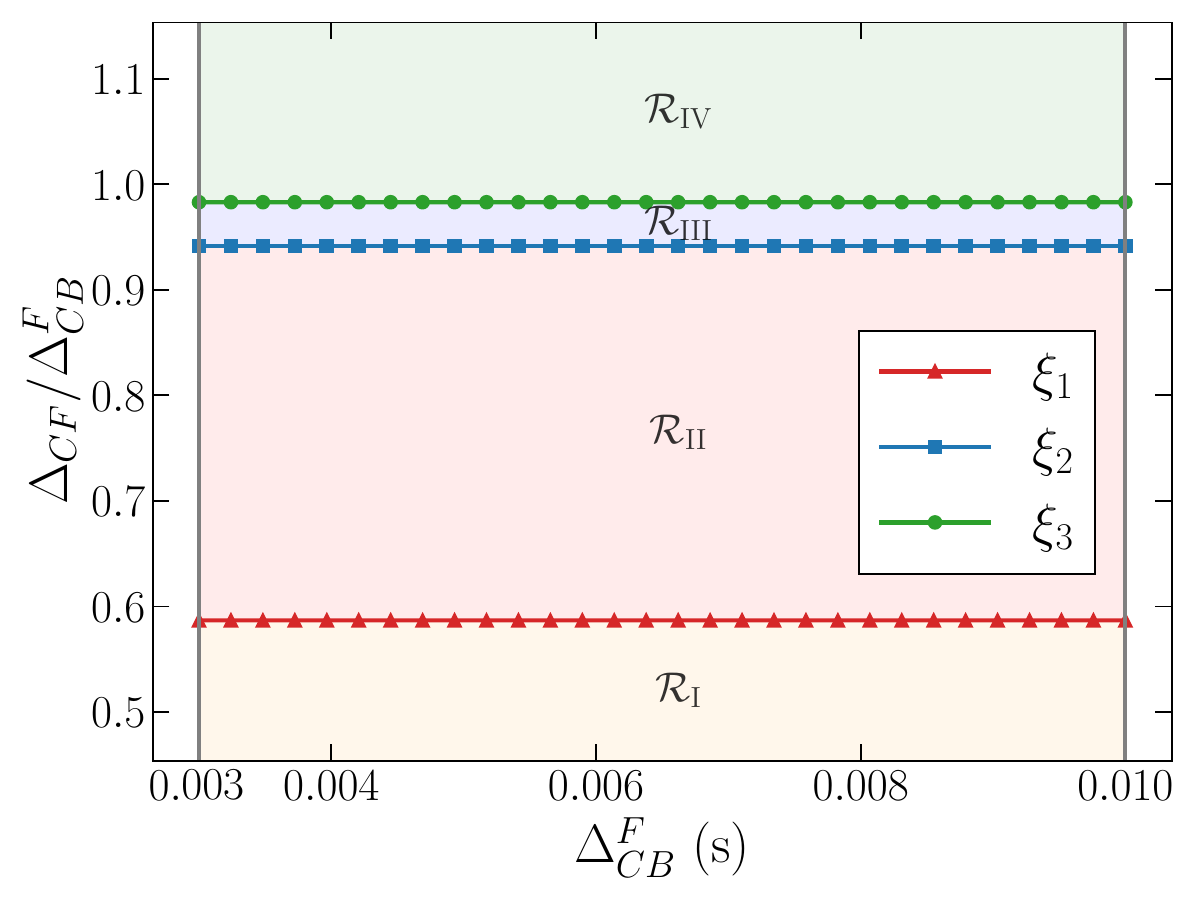}
        \end{minipage}
    }
    
    \caption{The variation of the three threshold ratios, $\xi_1, \xi_2,$ and $\xi_3$, and the corresponding four regions, $\mathcal{R}_\text{I}, \mathcal{R}_\text{II}, \mathcal{R}_\text{III},$ and $\mathcal{R}_\text{IV}$, with respect to the network parameters $n$, $\lambda_b$, $R$, and $\Delta_{CB}^F$. The default parameters are set as: $n=50$, $\lambda_b=10^2$ bit/s, $q=0.01$, $R=10^7$ bit/s, and $\Delta_{CB}^{F}=0.003$ s.}
    \label{regionsVarParams}
\end{figure*}

\begin{itemize}
    \item \textbf{Pronounced Advantage Region of the Connection-Free Scheme} $\mathcal{R}_\text{I} = \left\{ \Delta_{CF}/\Delta_{CB}^{F} \mid 0 < \Delta_{CF}/\Delta_{CB}^{F} < \xi_1 \right\}$. In this region, $\overline{T}_{CF}^*$ is less than $\overline{T}_{CB}^*$, and $\overline{T}_{CF}^{L=L_{CB}^*}$ is also less than $\overline{T}_{CB}^*$. This indicates that the connection-free scheme has a pronounced advantage in this region.

    \item \textbf{Large Packet Advantage Region of the Connection-Based Scheme} $\mathcal{R}_\text{II} = \left\{ \Delta_{CF}/\Delta_{CB}^{F} \mid \xi_1 < \Delta_{CF}/\Delta_{CB}^{F} < \xi_2 \right\}$. In this region, $\overline{T}_{CF}^*$ is less than $\overline{T}_{CB}^*$, but $\overline{T}_{CF}^{L=L_{CB}^*}$ is greater than $\overline{T}_{CB}^*$. The union $\mathcal{R}_\text{I} \cup \mathcal{R}_\text{II}$ constitutes the overall advantage region for the connection-free scheme, where it can achieve lower overall minimum delaycompared to the connection-based scheme. However, within $\mathcal{R}_\text{II}$ specifically, its delay performance at large packet sizes  is inferior to that of the connection-based scheme.

    \item \textbf{Small Packet Advantage Region of the Connection-Free Scheme} $\mathcal{R}_\text{III} = \left\{ \Delta_{CF}/\Delta_{CB}^{F} \mid \xi_2 < \Delta_{CF}/\Delta_{CB}^{F} < \xi_3 \right\}$. In this region, $\overline{T}_{CF}^*$ is greater than $\overline{T}_{CB}^*$, but $\overline{T}_{CF}^*$ is less than $\overline{T}_{CB}^{L=L_{CF}^*}$. This indicates that, although the connection-based scheme can achieve a lower overall minimum delaycompared to the connection-free scheme, its delay performance at small packet sizes  is inferior to that of the connection-free scheme.

    \item \textbf{Pronounced Advantage Region of the Connection-Based Scheme} $\mathcal{R}_\text{IV} = \left\{ \Delta_{CF}/\Delta_{CB}^{F} \mid \Delta_{CF}/\Delta_{CB}^{F} > \xi_3 \right\}$. In this region, $\overline{T}_{CF}^*$ is greater than $\overline{T}_{CB}^*$, and $\overline{T}_{CF}^*$ is also greater than $\overline{T}_{CB}^{L=L_{CF}^*}$. The union $\mathcal{R}_\text{III} \cup \mathcal{R}_\text{IV}$ forms the overall advantage region for the connection-based scheme, where it achieves lower overall minimum delaycompared to the connection-free scheme.
\end{itemize}

\subsection{Impact of Various Network Parameters on the Trade-off}
We now analyze the impact of various network parameters on the trade-off. Fig.~\ref{regionsVarParams} illustrates the variation of the three threshold ratios, $\xi_1, \xi_2,$ and $\xi_3$, and the corresponding four regions, $\mathcal{R}_\text{I}, \mathcal{R}_\text{II}, \mathcal{R}_\text{III},$ and $\mathcal{R}_\text{IV}$, with respect to the network parameters $n$, $\lambda_b$, $R$, and $\Delta_{CB}^F$.

As shown in Fig.~\ref{regions n}, as $n$ increases from 20 to 300, $\xi_3$ remains almost constant, while $\xi_1$ and $\xi_2$ exhibit a clear monotonic decrease. Consequently, the pronounced advantage region of the connection-free scheme $\mathcal{R}_\text{I}$, and its overall advantage region $\mathcal{R}_\text{I} \cup \mathcal{R}_\text{II}$, gradually shrink. Conversely, the overall advantage region for the connection-based scheme $\mathcal{R}_\text{III} \cup \mathcal{R}_\text{IV}$, and the large packet advantage region of the connection-based Scheme $\mathcal{R}_\text{II}$, progressively expand. Furthermore, an interesting observation is that the small packet advantage region of the connection-free scheme, $\mathcal{R}_\text{III}$, also gradually enlarges. These phenomena jointly indicate that in denser networks, the overall advantage and the large packet advantage of the connection-based scheme become more significant. However, even in such networks, the connection-free scheme can still be a feasible option for small packet transmission  for specific values of $\Delta_{CF}/\Delta_{CB}^F$.

Fig.~\ref{regions lambda_b} shows that the impact of $\lambda_b$ is similar to that of $n$. A notable phenomenon is that when $\lambda_b$ exceeds approximately $5\times 10^2$ bit/s, $\xi_1$ ceases to exist, and as a result, $\mathcal{R}_\text{I}$ disappears. This implies that when $\lambda_b$ is excessively large, the connection-free scheme can never achieve a pronounced advantage. For relatively large packets, the connection-based scheme will always be the superior choice, regardless of the $\Delta_{CF}/\Delta_{CB}^F$ value.

From Fig.~\ref{regions R}, it can be observed that when $R$ is below approximately $2 \times 10^6$ bit/s, $\xi_1$ does not exist and $\mathcal{R}_\text{I}$ vanishes, similar to the case of an excessively large $\lambda_b$. As $R$ increases from $10^5$ bit/s to $10^7$ bit/s, $\xi_1$ emerges and then monotonically increases, while $\xi_2$ and $\xi_3$ also increase monotonically. Correspondingly, the pronounced advantage region of the connection-free scheme $\mathcal{R}_\text{I}$, emerges and expands, and its overall advantage region $\mathcal{R}_\text{I} \cup \mathcal{R}_\text{II}$ also expands. In contrast, the overall advantage region for the connection-based scheme $\mathcal{R}_\text{III} \cup \mathcal{R}_\text{IV}$, and its pronounced advantage  region $\mathcal{R}_\text{IV}$, both shrink. The large packet advantage region of the connection-based scheme $\mathcal{R}_\text{II}$, first increases and then decreases. Similarly, it is interesting to note that the small packet advantage region of the connection-free scheme, $\mathcal{R}_\text{III}$, gradually shrinks. Collectively, these observations suggest that as $R$ increases, the pronounced advantage  and overall advantage of the connection-free scheme are enhanced, while all advantages of the connection-based scheme diminish. However, with an increasing $R$, it becomes more difficult for the connection-free scheme to leverage its advantage in small packet transmission  when the connection-based scheme is overall advantageous.

Since the preceding analyses in Fig.~\ref{fourRegions} and Figs.~\ref{regions n}–\ref{regions R} were conducted by holding $\Delta_{CB}^{F}$ constant while varying $\Delta_{CF}$, we further investigate the impact of different $\Delta_{CB}^{F}$ values in Fig.~\ref{regions deltaF}. As can be seen from the figure, as $\Delta_{CB}^{F}$ increases, the threshold ratios $\xi_1, \xi_2,$ and $\xi_3$, along with the corresponding four regions $\mathcal{R}_\text{I}, \mathcal{R}_\text{II}, \mathcal{R}_\text{III},$ and $\mathcal{R}_\text{IV}$, remain almost unchanged. This observation validates the generality of our previous analysis and reaffirms that the trade-off between the connection-free and connection-based schemes is governed by the ratio of their inherent core overheads, rather than their absolute values.

\section{Case Study: RA-SDT in NTN Scenarios}
\label{sectionV}
In this section, we apply the theoretical analysis to RA-SDT in NTN scenarios. First, we introduce the RA procedure in NTN scenarios. Subsequently, we investigate the scaling law relationship between the round trip time and both the optimal packet size and optimal delay. Finally, using NR TN as a baseline, we quantify the performance degradation in queueing delay and the variations in optimal packetization for NR NTN and IoT NTN scenarios under different values of the number of UEs $n$ and the bit arrival rate $\lambda_b$.

\subsection{RA Procedure and Modeling in NTN Scenarios}
The case study in \cite{10750858} employs suitable simplifications to characterize 2-step RA-SDT as a connection-free Aloha model and 4-step RA-SDT as a connection-based Aloha model. Our case study adopts the same simplification and characterization  proposed in  \cite{10750858}.

The analysis in \cite{10750858} implicitly assumes TN conditions, where the Round Trip Time (RTT) between the UE and gNB is considered negligible. However, for NTN, this assumption no longer holds, as the RTT becomes substantially larger. This significant propagation delay in NTN means that downlink transmissions from the gNB require a considerably longer time to reach the UE. A fundamental problem emerges if this RTT is not properly accounted for. In the 2-step procedure, a UE risks terminating its reception window for the MsgB response too early after a successful MsgA transmission. A similar problem affects the 4-step procedure, where the UE may stop attempting to detect Msg2 too soon after sending Msg1. This timing mismatch would force the UE to erroneously declare a transmission failure for its initial uplink message (MsgA or Msg1), leading to a persistent cycle of false failure detections and subsequent re-initiations of the random access procedure \cite{UEgNBERTT}.

To resolve this issue, 3GPP has approved the introduction of a time offset equal to the estimated RTT between the UE and gNB to the start of the downlink response window \cite{UEgNBERTT}. Specifically, for the 2-step RA-SDT, the UE delays by this RTT offset following its MsgA transmission before initiating the MsgB response window. This same principle applies to the 4-step procedure for the Msg2 and Msg4 response windows.

Furthermore, in the 4-step RA-SDT, a scheduling time offset, denoted as $K_{\mathrm{offset}}$, is introduced for the Msg3 transmission to avoid the uplink-downlink timing overlap caused by the large Timing Advance (TA) \cite{3gpp_R1_2007074, 3gpp_R1_2104099}. Since $K_{\mathrm{offset}}$ is designed to cover the overall propagation RTT between the UE and gNB, we set $K_{\mathrm{offset}} = \mathrm{RTT}_{\mathrm{UE-gNB}}$ in our modeling.

Adapting the case study parameters from \cite{10750858} and incorporating the aforementioned timing enhancement mechanisms, the time overhead parameters are determined as follows. For the 2-step RA-SDT, $\Delta_{CF}$ includes the delay compensation for the MsgB response window \cite{UEgNBERTT}, with the remaining constant overhead taken from \cite{10750858}:
\begin{equation}
\Delta_{CF} = \left(\mathrm{RTT}_{\mathrm{UE-gNB}} + 5.5\right) \times 10^{-3}.
\end{equation}

For the 4-step RA-SDT, $\Delta_{CB}^{F}$ includes the delay compensation for the Msg2 response window \cite{UEgNBERTT}, with the constant term again adopted from \cite{10750858}:
\begin{equation}
\Delta_{CB}^{F} = \left(\mathrm{RTT}_{\mathrm{UE-gNB}} + 2\right) \times 10^{-3}.
\end{equation}

Finally, for a successful transmission in the 4-step RA-SDT, $\Delta_{CB}^{S}$ comprises the delay compensation for the Msg2 and Msg4 response windows \cite{UEgNBERTT}, the $K_{\mathrm{offset}}$ scheduling offset for Msg3 \cite{3gpp_R1_2007074, 3gpp_R1_2104099}, and the remaining constant overhead from \cite{10750858}. It is calculated as:
\begin{equation}
\Delta_{CB}^{S} = \left(3 \times \mathrm{RTT}_{\mathrm{UE-gNB}} + 7.5\right) \times 10^{-3}.
\end{equation}

It is important to clarify that the above parameters represent a set of typical values for illustrative purposes \cite{10750858}, and they can vary in practical network deployments. Since our preceding analysis has established that the complex trade-off between the connection-free and connection-based schemes is critically determined by $\Delta_{CF}/\Delta_{CB}^{F}$, and this ratio is variable in practice, this case study does not aim to provide a general conclusion regarding the optimal queueing delay performance superiority of 2-step versus 4-step RA-SDT.

\begin{figure}[htbp]
    \centering
    \includegraphics[width=\columnwidth]{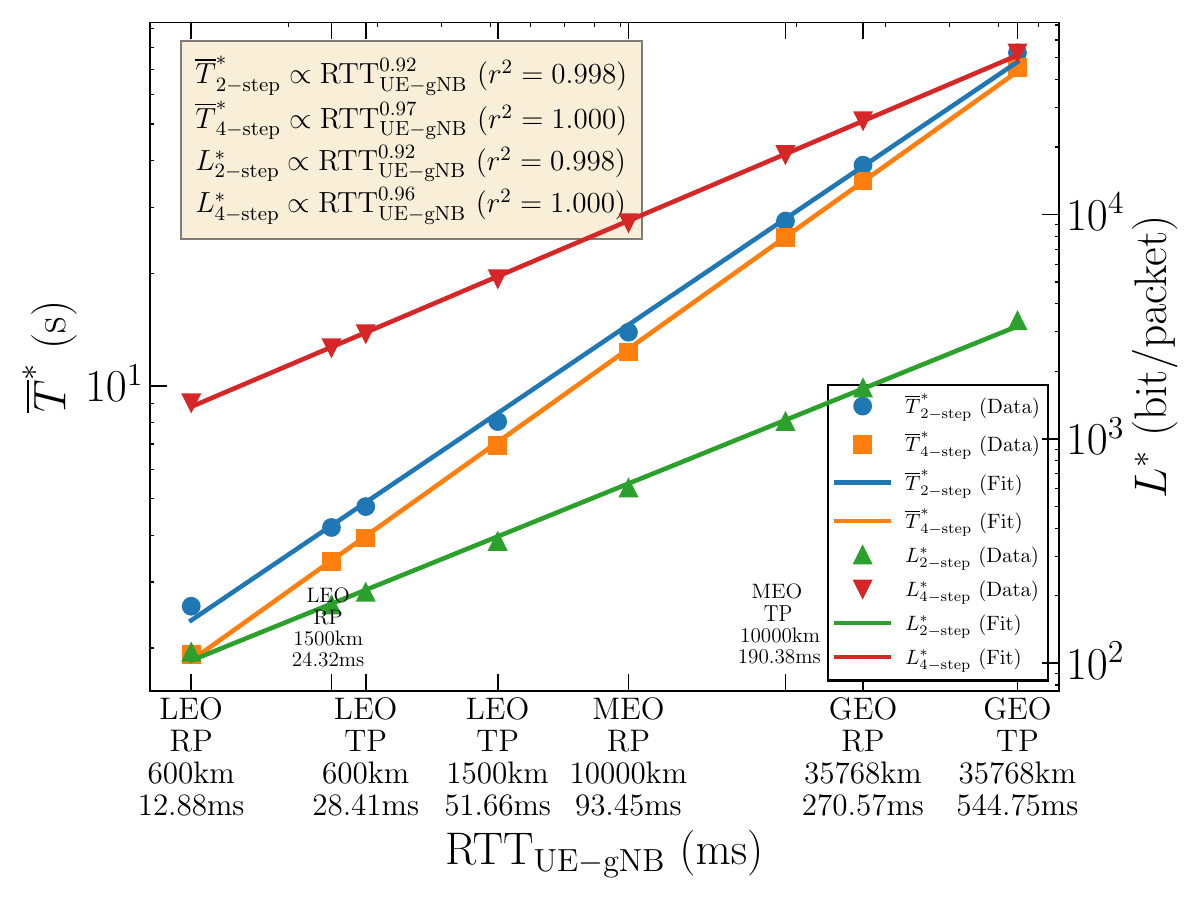}
    \caption{Calculated data points and regression fit curves for  optimal delay $\overline{T}^*$ and optimal packet size $L^*$ versus $\mathrm{RTT}_{\mathrm{UE-gNB}}$. $n=200$, $\lambda_b=1$bit/s, $q=0.008$, $R=10^5$bit/s.}
    \label{scaling law}
\end{figure}

\begin{figure}[htbp]
    \centering
    \includegraphics[width=0.9\columnwidth]{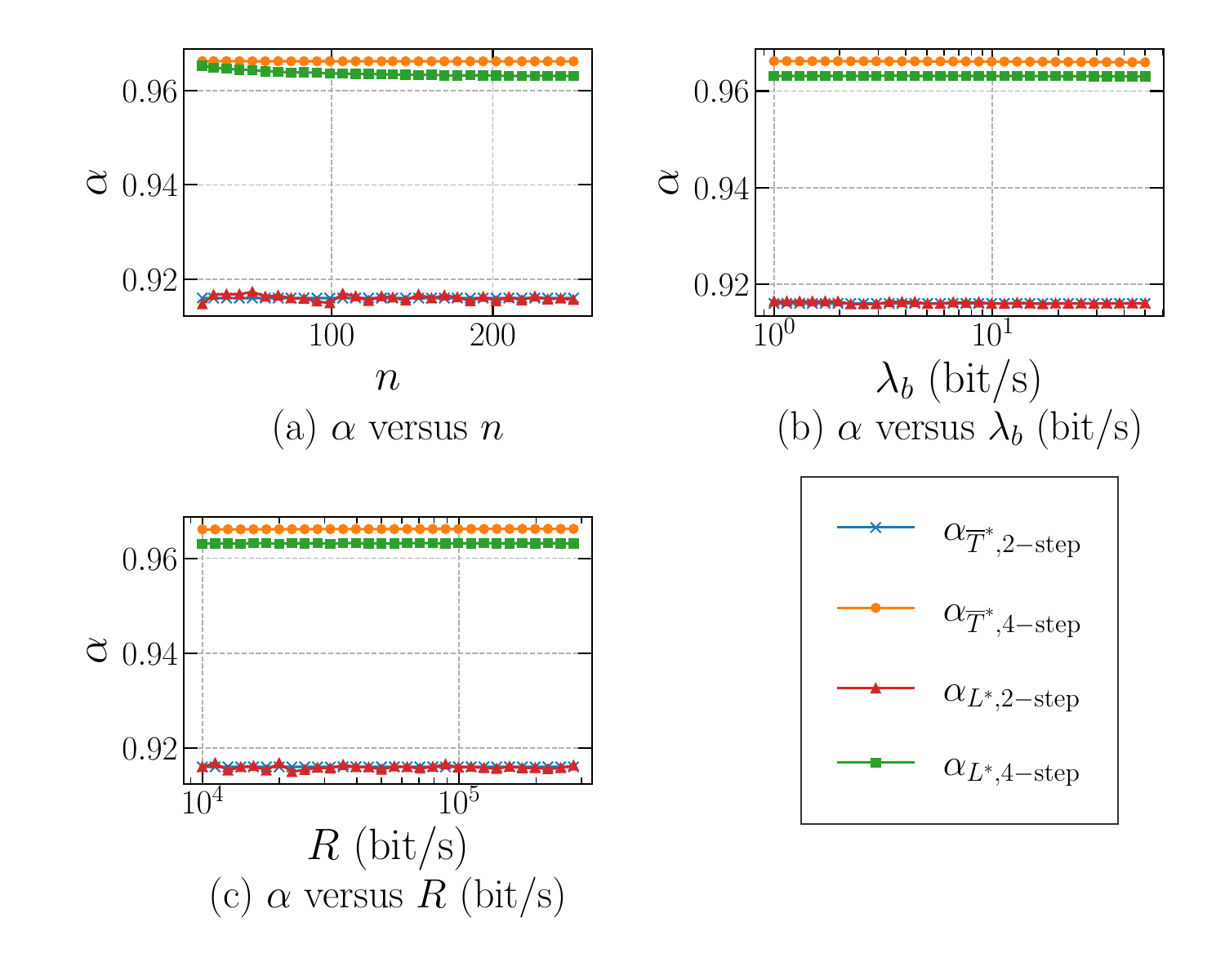}
    \caption{The scaling law exponent $\alpha$ versus various network parameters. The default parameters are set as: $n=200$, $\lambda_b=1$bit/s, $q=0.008$, $R=10^5$bit/s.}
    \label{alphaSensitivity}
\end{figure}

\begin{figure}[htbp]
    \centering
    \includegraphics[width=0.9\linewidth]{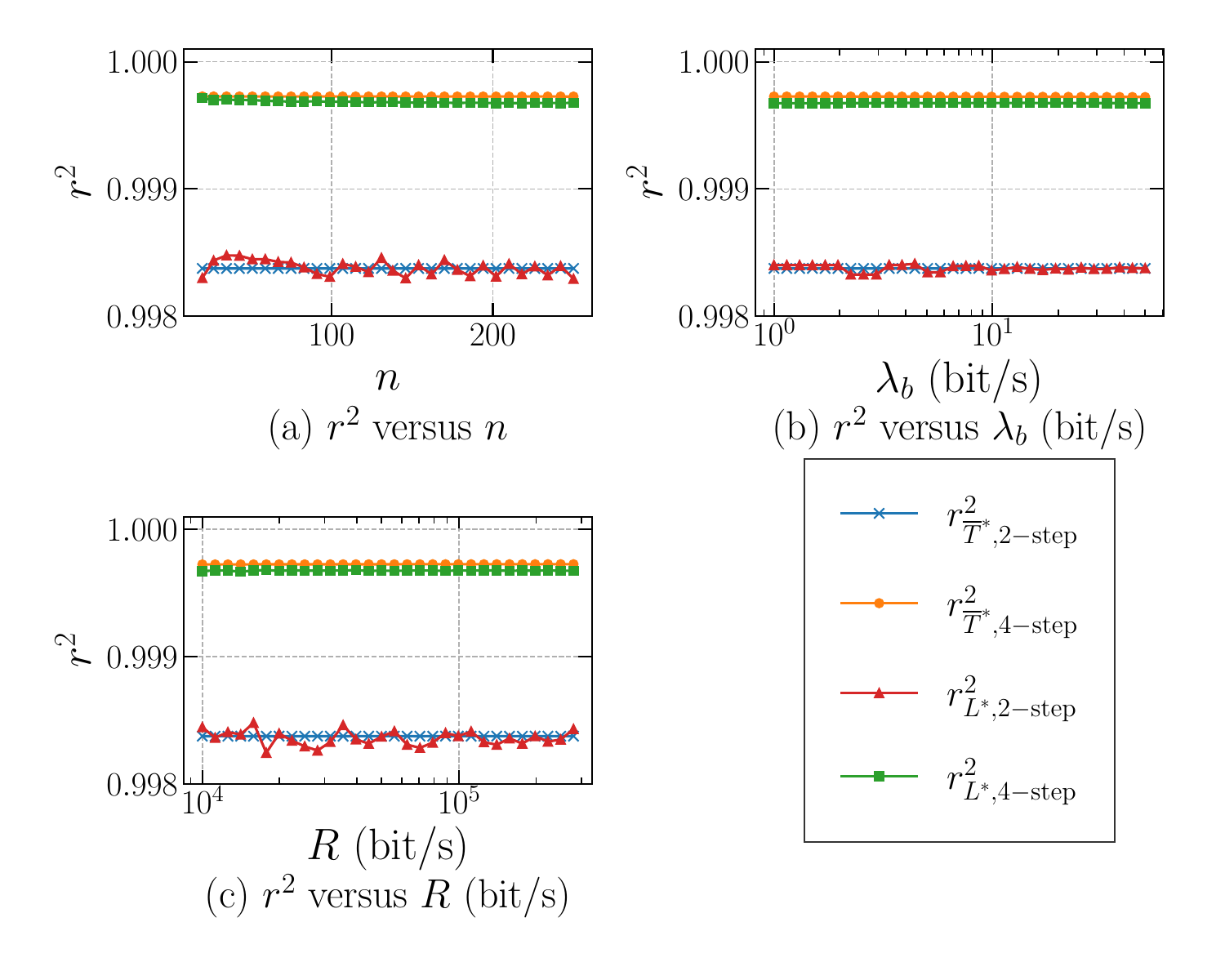}
    \caption{The coefficient of determination $r^2$ versus varying network parameters. The default parameters are set as: $n=200$, $\lambda_b=1$bit/s, $q=0.008$, $R=10^5$bit/s.}
    \label{r2Sensitivity}
\end{figure}

\subsection{Impact of RTT}

\begin{figure*}[htbp]
    \centering
    \subfigure[Relative delay versus number of UEs $n$.]{
        \label{relative delay n}
        \begin{minipage}{0.48\textwidth}
            \centering
            \includegraphics[width=\linewidth]{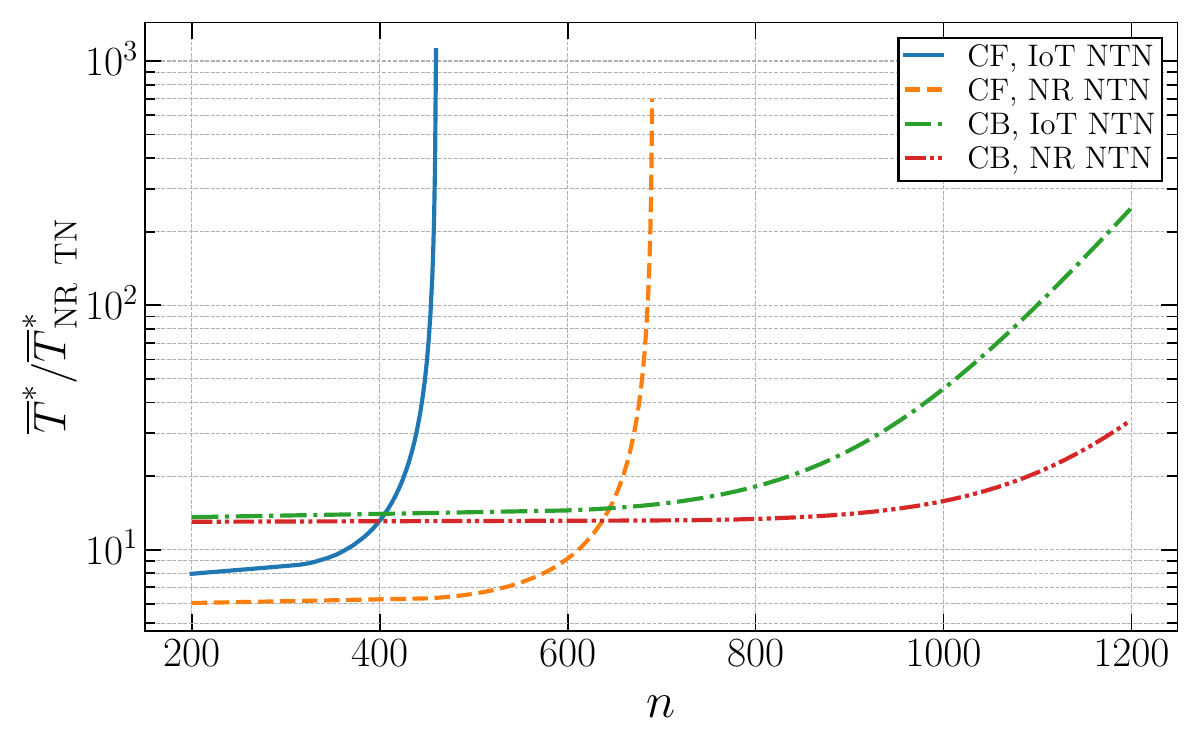}
        \end{minipage}
    }
    \hfill
    \subfigure[Relative delay versus bit arrival rate $\lambda_b$.]{
        \label{relative delay lambda}
        \begin{minipage}{0.48\textwidth}
            \centering
            \includegraphics[width=\linewidth]{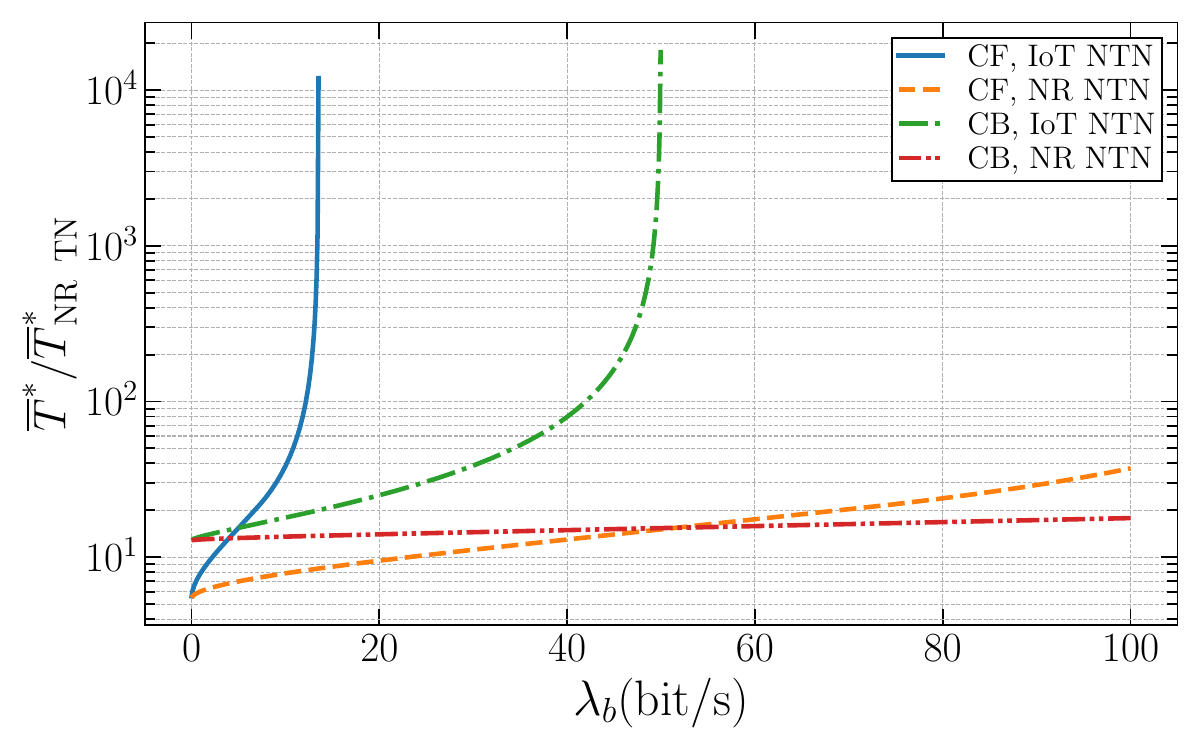}
        \end{minipage}
    }
    \\
    \vspace{1.5ex}
    \subfigure[Relative optimal packet size versus number of UEs $n$.]{
        \label{relative L n}
        \begin{minipage}{0.48\textwidth}
            \centering
            \includegraphics[width=\linewidth]{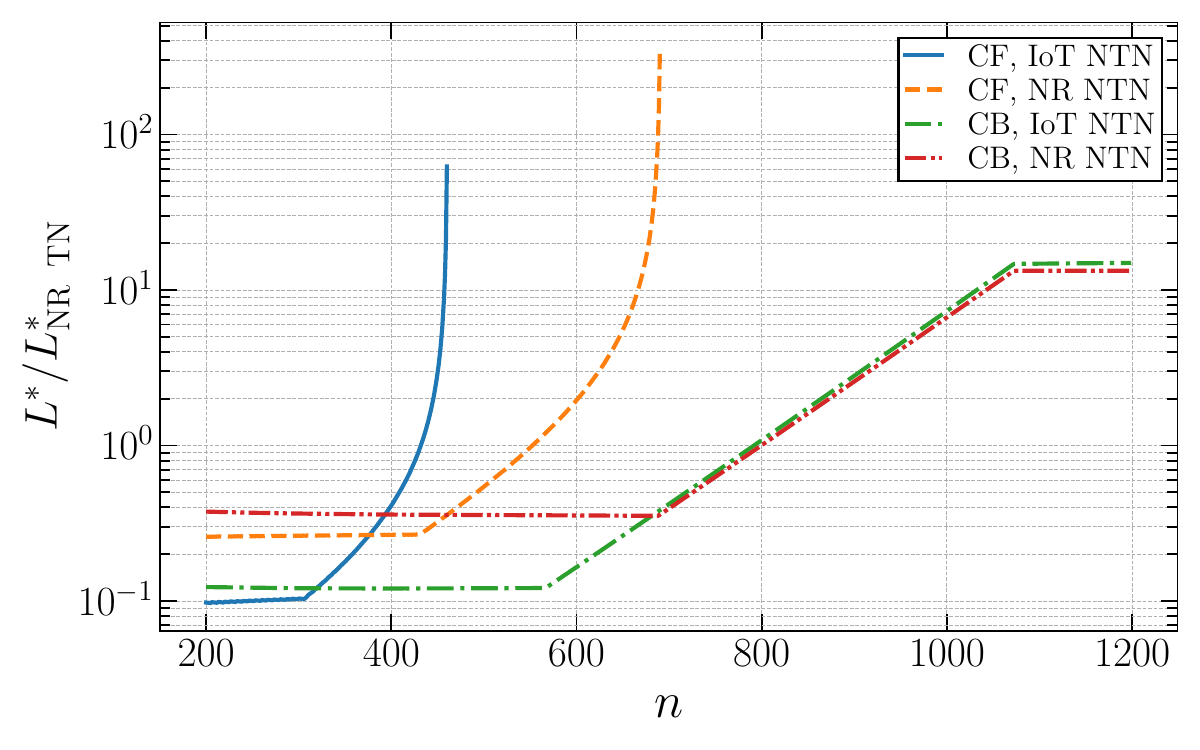}
        \end{minipage}
    }
    \hfill
    \subfigure[Relative optimal packet size versus bit arrival rate $\lambda_b$.]{
        \label{relative L lambda}
        \begin{minipage}{0.48\textwidth}
            \centering
            \includegraphics[width=\linewidth]{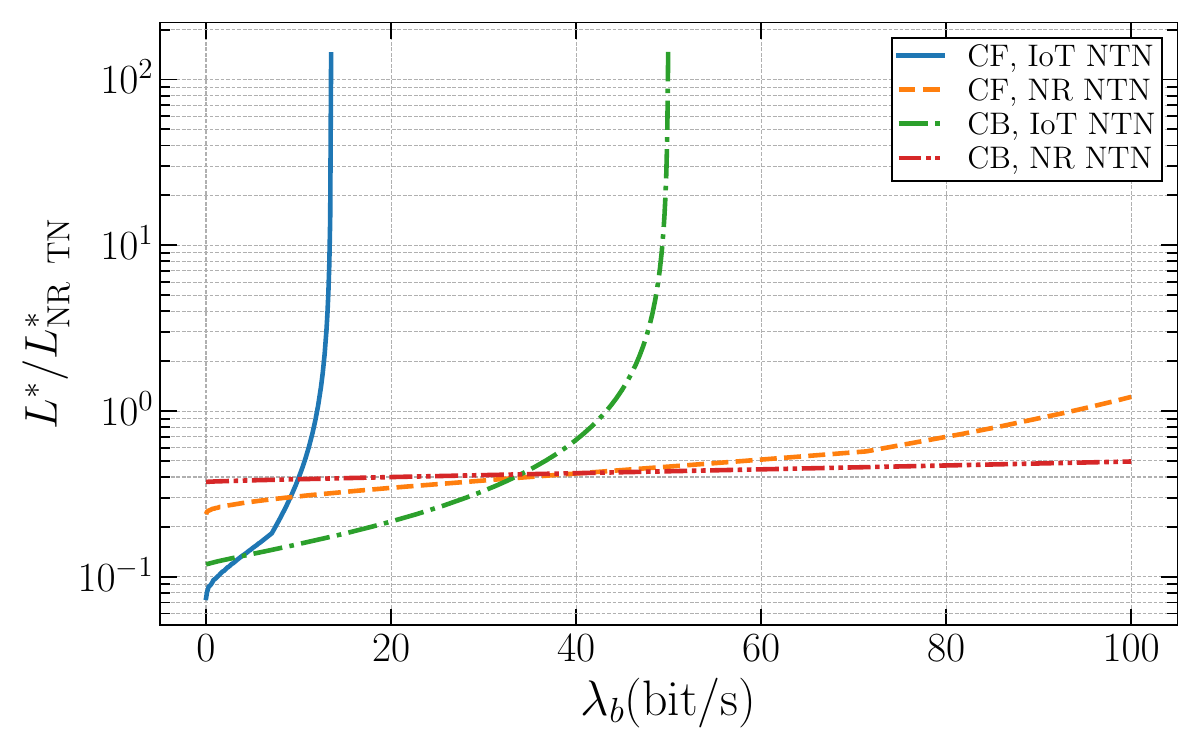}
        \end{minipage}
    }
    \caption{Relative optimal delay $\overline{T}^*/\overline{T}_\mathrm{NR\ TN}^*$ and relative optimal packet size $L^*/L_\mathrm{NR\ TN}^*$ for four different NTN scenarios under different values of number of UEs $n$ and bit arrival rate  $\lambda_b$. }
    \label{NTNComp}
\end{figure*}

\cite{3GPPRTT} provides the RTT between UE and gNB $\mathrm{RTT}_{\mathrm{UE-gNB}}$ for satellites at different altitudes with a UE elevation angle set at 10° and a ground gateway elevation angle at 5°, in both transparent payload (denoted as TP here) mode and regenerative payload (denoted as RP here) mode. Utilizing the data provided in \cite{3GPPRTT}, while neglecting changes in elevation angle caused by satellite motion, we observe a scaling law relationship between the $\mathrm{RTT}_{\mathrm{UE-gNB}}$ and both the optimal packet size and the minimum mean queueing delay. That is,
\begin{equation}
    \overline T^* = k\cdot (\mathrm{RTT}_{\mathrm{UE-gNB}})^{\alpha_{\overline T^*}},
\end{equation}
and
\begin{equation}
    L^* = k\cdot (\mathrm{RTT}_{\mathrm{UE-gNB}})^{\alpha_{L^*}},
\end{equation}
where $\alpha_{\overline T^*}$ and $\alpha_{L^*}$ are the  scaling law exponents for $\overline T^*$ and $L^*$, respectively.

As illustrated in Fig.~\ref{scaling law}, which uses a log-log scale for intuitive visualization, this underlying scaling law relationship is demonstrated through ordinary least squares linear regression. This scaling law relationship reveals a key property: when the $\mathrm{RTT}_{\mathrm{UE-gNB}}$ increases by a factor of $x$, the optimal delay $\overline{T}^*$ and optimal packet size $L^*$ increase by factors of approximately $x^{\alpha_{\overline T^*}}$ and $x^{\alpha_{L^*}}$, respectively. Moreover, the corresponding coefficient of determination $r^2$ for each of the four regression lines is close to 1, signifying an exceptionally high goodness of fit for the scaling law model.

Fig.~\ref{alphaSensitivity} and Fig.~\ref{r2Sensitivity} further investigate the influence of various network parameters on $\alpha$ and $r^2$. The results indicate that, within the wide parameter space explored, both $\alpha$ and $r^2$ maintain excellent stability. This stability strongly demonstrates the general nature of this scaling law relationship.

\subsection{Performance Comparison of NTN and TN Scenarios}
We now quantify the performance degradation in queueing delay and the corresponding variations in optimal packetization for different NTN scenarios. The performance of each NTN scheme is normalized by that of its corresponding mode in a single terrestrial baseline scenario: a 5G New Radio Terrestrial Network (NR TN). That is, both 2-step NTN schemes (NR and IoT) are compared against the 2-step NR TN baseline, and both 4-step NTN schemes are compared against the 4-step NR TN baseline. The results are shown in Fig.~\ref{NTNComp}, and the parameters for each scenario are listed in Table~\ref{caseStudyParams}. The values of $R$ for the three scenarios are obtained from the user-experienced data rate requirements mentioned in \cite{3GPPR, 5GR, IoTR}. For both the NR NTN and IoT NTN scenarios, the $\mathrm{RTT}_{\mathrm{UE-gNB}}$ is set as 24.32 ms.

\begin{table}[htbp]
    \renewcommand{\arraystretch}{1.2}
    \centering
    \caption{Parameters Employed in the Evaluation}
    \begin{tabular}{lccc}
        \toprule
        Parameter & NR NTN & IoT NTN & NR TN \\
        \midrule
        $R$ (bit/s) & $10^5$ & $10^4$ & $5 \times 10^7$ \\
        $\mathrm{RTT}_{\mathrm{UE-gNB}}$ (ms) & $24.32$ & $24.32$ & $0$ \\
        $q$ & 0.01 & 0.01 & 0.01 \\
        \bottomrule
    \end{tabular}
    \label{caseStudyParams}
\end{table}

As shown in Fig.~\ref{relative delay n} and Fig.~\ref{relative L n}, the relative optimal delay and relative optimal packet size exhibit similar trends as $n$ increases. For all four NTN scenarios, there is an initial range of $n$ where both performance metrics remain relatively stable. For instance, these stable regions for the 2-step IoT NTN, 2-step NR NTN, 4-step IoT NTN, and 4-step NR NTN schemes extend up to approximately $n=300$, $n=420$, $n=580$, and $n=700$, respectively. Beyond these ranges, the relative delay and optimal packet size for the 2-step RA-SDT schemes increase sharply as the network rapidly approaches saturation. In contrast, the 4-step RA-SDT schemes exhibit a much more smooth increase, which confirms the aforementioned property that connection-based schemes are less sensitive to $n$. A comparison between the IoT and NR scenarios shows that the IoT NTN schemes, which operate with a lower data rate $R$, experience more severe delay degradation and are more sensitive to $n$, with a smaller stable range. These observations suggest that, compared to TN, the number of UEs in NTN RA-SDT must be strictly controlled to avoid severe performance degradation, particularly for 2-step and IoT NTN schemes. Furthermore, an interesting phenomenon is observed for the optimal packetization: for small values of $n$, the relative optimal packet size for all four NTN schemes is less than one, indicating a smaller packet size compared to the terrestrial baseline. As $n$ increases, however, the optimal packet sizes surpass the baseline and become significantly larger. Another notable trend is that for the 4-step schemes, the relative packet size appears to enter another stable region once $n$ exceeds about 1050. The above phenomenon highlights the need to carefully select the optimal packetization based on the number of UEs.

Since the focus is on small data transmission, we limit the considered range of $\lambda_b$ to 100 bit/s. As can be seen from Fig.~\ref{relative delay lambda} and Fig.~\ref{relative L lambda}, the relative optimal delay and the relative optimal packet size exhibit similar trends as $\lambda_b$ increases. The IoT schemes, both 2-step and 4-step, exhibit a high sensitivity to $\lambda_b$, with both their relative delay and optimal packet size increasing rapidly from $\lambda_b=0$. The 2-step IoT NTN network approaches saturation beyond $\lambda_b \approx 10$ bit/s, while the 4-step IoT NTN network does so beyond $\lambda_b \approx 50$ bit/s. In stark contrast, the NR schemes show strong stability, especially the 4-step NR NTN, for which both the relative delay and optimal packet size remain almost horizontal across the considered range of $\lambda_b$. In summary, in IoT NTN scenarios, the bit arrival rate $\lambda_b$ must be strictly controlled for both 2-step and 4-step schemes to prevent severe performance degradation. Furthermore, the substantial variation in relative optimal packet size  across different $\lambda_b$ values underscores the critical importance of carefully determining the optimal packet size of 2-step schemes based on the specific $\lambda_b$ conditions.

\section{Conclusion}
\label{sectionVI}
In this paper, we conducted an in-depth investigation into the relationship between packetization and queueing delay measured in seconds, which revealed a series of new findings from this perspective.
We found that for both connection-free and connection-based schemes, the optimal packetization varies with the transmission probability $q$. For smaller values of $q$, an optimal packet size $L^*$ exists at a delay extremum within the unsaturated operational range; for larger values of $q$, the optimal packet size  equals the minimum packet size that keeps the network unsaturated. Our investigation into the impact of various network parameters revealed the different sensitivities of the two schemes. We showed that the connection-based scheme exhibits stronger robustness against increases in $n$ and $\lambda_b$, while the connection-free scheme benefits more from a high $R$. Our analysis also identified $\Delta_{CF}$ and $\Delta_{CB}^{F}$ as the most influential factors for the queueing delay performance of connection-free and connection-based schemes, respectively. Simulations of jitter demonstrated a similar relationship with packetization, indicating a strong synergy between mean queueing delay and jitter of queueing delay performance for the connection-free scheme, and a degree of synergy for the connection-based scheme as well.

Our analysis was then applied to re-evaluate the trade-off between the two schemes from the perspective of packetization. By characterizing three distinct thresholds $\xi_1$ \string~  $\xi_3$ based on the overhead ratio $\Delta_{CF}/\Delta_{CB}^{F}$, which divide the parameter space into four different regions  $\mathcal{R}_\text {I}$ \string~  $\mathcal{R}_\text {IV}$, we described the different advantages of each scheme. Furthermore, by exploring how these boundaries vary with network parameters, we comprehensively characterized their dynamics.

Finally, as a case study, we applied our analysis to RA-SDT in NTN scenarios. We identified a scaling law relationship between the round trip time and both the optimal packet size and optimal delay. Our findings also showed that, compared to TN, the number of UEs in NTN RA-SDT must be strictly controlled to avoid severe performance degradation, particularly for 2-step and IoT NTN schemes. In IoT NTN scenarios, the bit arrival rate must be strictly controlled for both 2-step and 4-step schemes.

\bibliographystyle{IEEEtran}
\bibliography{IEEEabrv,reference}
\end{document}